\batchmode
\makeatletter
\def\input@path{{/Users/dill/current/varprop/}}
\makeatother
\documentclass[12pt,british,english]{article}\usepackage[]{graphicx}\usepackage[]{color}
\makeatletter
\def\maxwidth{ %
  \ifdim\Gin@nat@width>\linewidth
    \linewidth
  \else
    \Gin@nat@width
  \fi
}
\makeatother

\definecolor{fgcolor}{rgb}{0.345, 0.345, 0.345}
\newcommand{\hlnum}[1]{\textcolor[rgb]{0.686,0.059,0.569}{#1}}%
\newcommand{\hlstr}[1]{\textcolor[rgb]{0.192,0.494,0.8}{#1}}%
\newcommand{\hlcom}[1]{\textcolor[rgb]{0.678,0.584,0.686}{\textit{#1}}}%
\newcommand{\hlopt}[1]{\textcolor[rgb]{0,0,0}{#1}}%
\newcommand{\hlstd}[1]{\textcolor[rgb]{0.345,0.345,0.345}{#1}}%
\newcommand{\hlkwa}[1]{\textcolor[rgb]{0.161,0.373,0.58}{\textbf{#1}}}%
\newcommand{\hlkwb}[1]{\textcolor[rgb]{0.69,0.353,0.396}{#1}}%
\newcommand{\hlkwc}[1]{\textcolor[rgb]{0.333,0.667,0.333}{#1}}%
\newcommand{\hlkwd}[1]{\textcolor[rgb]{0.737,0.353,0.396}{\textbf{#1}}}%

\usepackage{framed}
\makeatletter
 {\par\unskip\endMakeFramed%
 \at@end@of@kframe}
\makeatother

\definecolor{shadecolor}{rgb}{.97, .97, .97}
\definecolor{messagecolor}{rgb}{0, 0, 0}
\definecolor{warningcolor}{rgb}{1, 0, 1}
\definecolor{errorcolor}{rgb}{1, 0, 0}
\newenvironment{knitrout}{}{} 

\usepackage{alltt}
\usepackage{lmodern}

\usepackage[T1]{fontenc}
\usepackage[utf8]{inputenc}
\usepackage[a4paper]{geometry}
\usepackage{bm}
\usepackage{amsmath}
\usepackage{amsthm}
\usepackage{amssymb}
\usepackage[authoryear]{natbib}

\makeatletter
\newcommand{\lyxaddress}[1]{
	\par {\raggedright #1
	\vspace{1.4em}
	\noindent\par}
}

\date{ }
\setcitestyle{round}

\makeatother

\usepackage{babel}
\IfFileExists{upquote.sty}{\usepackage{upquote}}{}
\begin{document}
\title{Variance propagation for density surface models}
\author{Mark V. Bravington \textsuperscript{1{*}}\\
David L. Miller \textsuperscript{2}\thanks{Joint first author. \texttt{dave@ninepointeightone.net}}\\
Sharon L. Hedley}
\maketitle

\lyxaddress{Commonwealth Scientific and Industrial Research Organisation Marine
Laboratory, Hobart, Australia\textsuperscript{1} and Centre for Research
into Ecological and Environmental Modelling and School of Mathematics
and Statistics, University of St Andrews, St Andrews, Fife, Scotland\textsuperscript{2}}
\begin{abstract}
Spatially-explicit estimates of population density, together with
appropriate estimates of uncertainty, are required in many management
contexts. Density Surface Models (DSMs) are a two-stage approach for
estimating spatially-varying density from distance-sampling data.
First, detection probabilities---perhaps depending on covariates---are
estimated based on details of individual encounters; next, local densities
are estimated using a GAM, by fitting local encounter rates to location
and/or spatially-varying covariates while allowing for the estimated
detectabilities. One criticism of DSMs has been that uncertainty from
the two stages is not usually propagated correctly into the final
variance estimates. We show how to reformulate a DSM so that the uncertainty
in detection probability from the distance sampling stage (regardless
of its complexity) is captured as an extra random effect in the GAM
stage. In effect, we refit an approximation to the detection function
model at the same time as fitting the spatial model. This allows straightforward
computation of the overall variance via exactly the same software
already needed to fit the GAM. A further extension allows for spatial
variation in group size, which can be an important covariate for detectability
as well as directly affecting abundance. We illustrate these models
using point transect survey data of Island Scrub-Jays on Santa Cruz
Island, CA and harbour porpoise from the SCANS-II line transect survey
of European waters.

\textbf{Keywords:} abundance estimation; distance sampling; generalized
additive models; line transect sampling; point transect sampling;
spatial modelling; 
\end{abstract}

\section{Introduction}

\setlength{\abovedisplayskip}{8pt}
\setlength{\belowdisplayskip}{8pt}

Distance sampling is a widely-used method for estimating abundance
when detection is imperfect \citep{Buckland:2001vm}, based on encounters
along line or point transects. Detection probability (detectability)
is estimated using within-encounter data (e.g., perpendicular distance
from trackline), by fitting ``detection functions'' that may involve
environmental covariates (e.g., local weather conditions). In traditional
stratified distance-sampling, an average animal density is then estimated
within each survey stratum---i.e., some region within which survey
coverage is supposed to be uniform---based on the observed encounter
rate within that stratum divided by the detectability, and then scaled
by the stratum area. Since the abundance estimate is a simple function
of statistically independent quantities (encounter rate and detectability),
its variance can be estimated straightforwardly. 

Instead of using strata, with modern statistical tools it is possible
to fit spatially-explicit models of density, where local density is
assumed to vary gradually in space (and perhaps also in response to
specific environmental covariates, which we here include under the
general heading of ``spatially-explicit''). Spatially-explicit estimates
are advantageous in many situations: when abundance estimates are
required across arbitrary sub-regions that do not coincide with survey
strata; to reduce bias when coverage is uneven; or when identifying
particularly important habitat for conservation, for example. 

There are various approaches to actually fitting spatially-explicit
models. The general idea, as in the stratified case, is that the expected
local encounter rate is the product of local detectability and local
density, but with both factors now potentially depending on local
spatial and/or enviromental covariates. Here we consider specifically
Density Surface Models \citep[DSMs;][]{Hedley:2004et,Miller:2013fq},
which take a two-stage approach. The first stage is to estimate detectability
using a detection function model; any standard or bespoke model could
be used (see Section \ref{sec:Density-surface-models}). In the second
stage, the encounter rate data are fitted to location/environmental
covariates using a GAM, specifically the ``basis-and-penalty'' formulation
of GAMs in \citet{wood2017generalized} in which smoothers are represented
via random effects. The estimated detectabilities for each segment
of search effort are easily accommodated in the GAM (technically,
as offsets to the linear predictor; see below), and the range of smoothers
and interactions that can be fitted in is very wide. 

Splitting the analysis into two stages is appealing partly because
existing domain-specific software and diagnostic expertise can be
applied as-is to each stage separately, and partly because it avoids
any need to write inevitably complicated code that incorporates two
individually-complex aspects. It is also straightforward to produce
a point estimate of abundance for any desired sub-region straight
from the fitted GAM. However, when detectability and density both
vary spatially, the problem is what to do about variance given that
GAMs do not intrinsically ``understand'' the notion of uncertainty
in their offset. 

In this paper, we show how statistical uncertainty about detectability
can in fact be accommodated painlessly within standard GAM software.
Our approach is to first fit the detection function as usual, but
then to rewrite the fitted detection function log-likelihood as a
quadratic approximation centred on its point estimates, and to incorporate
the uncertainty about the detection function parameters via random
effects in the second-stage GAM. This fits directly and automatically
into the Wood/Wahba formulation of a GAM, whereby a smooth surface
is described by a set of coefficients treated as random effects; thus,
the machinery for handling random effects in general is already built
into the \texttt{mgcv} software used in \citet{Miller:2013fq}'s DSM
code. This amounts to re-fitting the detection function model (or
a good approximation to it) at the same time as fitting the GAM. We
retain the benefits of two-stage modeling, but all the uncertainty
about detectability as well as density is now captured in the usual
GAM outputs. The re-fitted detection function model should not differ
greatly from the original fit, and we use this idea to propose a diagnostic
for overall model specification issues. 

Following a summary of DSMs and notation in Section \ref{sec:Density-surface-models},
we present our new formulation in Section \ref{sec:varprop}, including
variance computation and diagnostics. Section \ref{sec:prev} comments
on problems with existing approaches to variance propagation in DSMs.
In Section \ref{sec:group-size-model}, we extend the formulation
to cover DSMs where group size varies spatially and affects detectability
(a common situation with whales and dolphins). Section \ref{examples}
gives examples of the variance propagation method and the group size
model. Some discussion is given in Section \ref{sec:Discussion},
including possible generalizations. 

\section{Density surface models\label{sec:Density-surface-models}}

In distance sampling observers move along a set of survey lines or
between points, counting (groups of) animals, recording distances
from the centre line or centre point to the observed groups (or their
cues, such as blows for cetaceans or calls for birds), the size of
each detected group and potentially other covariates that may effect
detectability. 

To fully describe the DSMs in this paper, we distinguish four different
classes of variable.
\begin{enumerate}
\item Density covariates, $x$, vary in space and potentially affect local
animal abundance: e.g., latitude and depth. They are required for
prediction and fitting, and are assumed known across the entire region
of interest.
\item Effort covariate(s), $z$, affect detection probability: e.g., sea
conditions measured on the Beaufort scale, or observer identity. They
are assumed known along each transect, but not necessarily in unsurveyed
areas.
\item Individual covariates, $g$, that affect detection probability and
are a persistent property of each group (independent of whether the
group is observed or not) during its window of observability: e.g.,
size (number of animals), and perhaps \foreignlanguage{british}{behaviour}.
Here $g$ is assumed known for each observed group (see Discussion).
The random variable $G$ varies from one group to the next, and its
statistical distribution $F_{G}\left(g;x\right)$ may vary spatially.
$F_{G}\left(g;x\right)$ may have a direct effect on abundance (via
the mean group size), as well as on detection probability, in which
case it is also necessary to estimate certain properties of $F_{G}$$\left(g;x\right)$
such as its local mean. 
\item Observation variables, $y$, which are random properties of one observation
on one group: e.g., perpendicular distance between the group and the
sampler. In certain settings, $y$ may contain other elements. For
example, in a multi-observer-platform survey \citep[e.g., MRDS;][]{Borchers:1998eq},
$y$ might also include which of the active observers saw the group;
in a cue-based setting, $y$ might include the bearing between sighting
and observer.
\end{enumerate}
These classes are assumed to be mutually exclusive; overlaps can lead
to fundamental problems for distance sampling which we do not address
here \citep[e.g., non-uniform animal distribution within the sample unit;][]{Marques:2012fy}.
The distinction between \emph{individual} and \emph{effort} covariates
is often glossed over but they have rather different implications
for abundance estimation (see below).

In the first stage of DSM, the detection function $\pi\left(y|\theta,z,g\right)$,
which involves unknown parameters $\theta$ as well as $z$ and $g$,
describes the probability of making an observation at $y$. The parameters
$\theta$ are usually estimated by maximizing this log-likelihood
across observations $s$:

\begin{gather}
l\left(\boldsymbol{\theta}\right)=\sum_{s}\log_{e}\left(\frac{\pi\left(y_{s}|\boldsymbol{\theta},z_{t_{s}},g_{s}\right)}{p\left(\boldsymbol{\theta};z_{t_{s}},g_{s}\right)}\right)\label{eq:detfct-ll}
\end{gather}
where $t_{s}$ is the transect containing sighting $s$. Here $p$
is the overall detection probability for a group, defined by
\[
p\left(\boldsymbol{\theta};z,g\right)=\int\pi\left(y;\boldsymbol{\theta},z,g\right)dF_{Y}\left(y\right)
\]
where $F_{Y}$ is the distribution function of $y$. In standard distance
sampling where $y$ consists only of perpendicular distance, $F_{Y}$
is uniform between $0$ and some fixed truncation distance, beyond
which observations are discarded. This formulation encompasses a wide
range of models, including multiple covariate distance sampling \citep[MCDS;][]{Marques:2003vb}
with $z$ and $g$, multi-observer mark-recapture distance sampling
\citep[MRDS;][]{Borchers:1998eq}, and cue-based ``hazard probability
models'' \citep{Skaug:1999vb}.

The second part of DSM models the local count of observations via
a GAM to capture spatial variation in animal density. This allows
us both to estimate abundance within any sub-region of interest, and
to compensate as far as possible for uneven survey coverage (whether
by design, or by virtue of field logistics and weather conditions).
Since line transects are generally very long in comparison to their
width and therefore contain a range of density and density covariate
values, we divide transects into smaller \textit{\emph{segment}}s,
which are the sample units for GAM (in which case the subscript $t_{s}$
above refers to segments rather than transects). Point transects are
left as-is; we use the term ``segments'' from now on to refer to
both points and line segments, without loss of generality. Environmental
covariates are assumed not to change much within each segment. The
relationship between counts $n_{i}$ per segment $i$ and density
covariates $x_{ik}$ is modelled as an additive combination of smooth
functions with a $\log$ link:

\begin{gather}
\mathbb{E}\left[n_{i}|\boldsymbol{\beta},\boldsymbol{\lambda},p(\hat{\boldsymbol{\theta}};\mathbf{z}_{i})\right]=a_{i}p(\hat{\boldsymbol{\theta}};\mathbf{z}_{i})\exp\left(\beta_{0}+\sum_{K}f_{k}(x_{ik})\right),\label{dsm-eqn}
\end{gather}
where each segment is of area $a_{i}$, and $n_{i}$ follows some
count distribution such as quasi-Poisson, Tweedie, or negative binomial.
The $f_{k}$ are smooth functions, represented by a basis expansion
($f_{k}(x)=\sum_{j}\beta_{j}b_{j}(x)$, for some basis functions $b_{j}$);
$\beta_{0}$ is an intercept term, included in parameter vector $\boldsymbol{\beta}$;
$\boldsymbol{\lambda}$ is a vector of smoothing (hyper)parameters
which control the wiggliness of the$f_{k}$. We take a Bayesian interpretation
of GAMs, in which $\boldsymbol{\lambda}$ controls the variance of
a multivariate improper Gaussian prior \citep{wood2017generalized}:
\[
\boldsymbol{\beta}\sim N\left(\mathbf{0},\phi\left(\sum_{k}\lambda_{k}\mathbf{S}_{k}\right)^{-}\right),
\]
with scale parameter $\phi$ , smoothing parameters $\lambda_{k}$
and penalty matrices $\mathbf{S}_{k}$ ($^{-}$ indicates pseudoinverse).
This leads to a quadratic penalty on beta during fitting. We estimate
$\boldsymbol{\lambda}$ itself via REML \citep{Wood:2011cl}, an empirical
Bayes procedure. Fully Bayesian approaches, placing hyperpriors on
$\boldsymbol{\lambda}$ are also possible.

We are interested in the uncertainty of a predicted abundance estimate,
$\hat{N}$. We assume below that we have created some prediction grid
with all density covariates available for each cell in the grid. Abundance
is predicted for each cell, and summed for an overall abundance, $\hat{N}$,
over some region of interest which may not be the entire surveyed
area. Although $p(\hat{\boldsymbol{\theta}})$ does not appear explicitly
in the prediction, which is
\[
\hat{N}=\sum_{j}a_{j}\exp\left(\hat{\beta_{0}}+\sum_{k}\hat{f_{k}}(x_{jk})\right),
\]
the GAM offsets $p(\hat{\boldsymbol{\theta}})$ clearly do affect
$\hat{\boldsymbol{\beta}}$, so it is important to account somehow
for detection probability uncertainty. (\ref{dsm-eqn}) assumes the
offset is fixed, so extra steps are required.

\section{Variance propagation for Density Surface Models}

\label{sec:varprop}Let $p\left(\boldsymbol{\theta}_{0},z_{i}\right)$
be the true probability of detection in segment $i$ and for now omit
$g$, therefore assuming that there are no individual-level covariates
(e.g., that group size is always 1) for now (see Section \ref{sec:group-size-model}).
If $\boldsymbol{\theta}_{0}$ is the true (unknown) value of $\boldsymbol{\theta}$,
and $\hat{\boldsymbol{\theta}}$ is its MLE, we use the shorthand
$p_{i}=p\left(\boldsymbol{\theta}_{0},z_{i}\right)$ and $\hat{p}_{i}=p\left(\hat{\boldsymbol{\theta}},z_{i}\right)$
when the dependence is clear. The expected number of encounters in
segment $i$ is $a_{i}p_{i}\rho_{i}$ where $\rho_{i}$ is the underlying
density, given by the exponential term in \ref{dsm-eqn}.

Given $p_{i}$, we can re-write (\ref{dsm-eqn}) on the $\log$ link
scale as:
\begin{gather}
\log\mathbb{E}\left[n_{i}|\boldsymbol{\beta},\boldsymbol{\lambda},p_{i}\right]=\eta_{i}=\log a_{i}p_{i}+X_{i}\boldsymbol{\beta}.\label{eq:log-link}
\end{gather}
$X_{i}$ is the (known) $i$\textsuperscript{th} row of the design
matrix, i.e., the values of the basis functions in segment $i$, so
$\log\rho_{i}=\sum_{k}f_{k}(x_{ik})=X_{i}\boldsymbol{\beta}$ and
$\log a_{i}p_{i}$ is an offset. The complication is that we only
have an estimate of $p_{i}$. To tackle this, we first rewrite the
linear predictor $\eta_{i}$ as

\begin{gather*}
\eta_{i}=\log a_{i}+\log\hat{p}_{i}+\log p_{i}-\log\hat{p}_{i}+X_{i}\boldsymbol{\beta}
\end{gather*}
and then take a Taylor series expansion of $\log\hat{p}_{i}\equiv\log p\left(\hat{\boldsymbol{\theta}},z_{i}\right)$
about $\boldsymbol{\theta}=\boldsymbol{\theta_{0}}$:

\begin{gather}
\log p\left(\hat{\boldsymbol{\theta}},z_{i}\right)=\log p\left(\boldsymbol{\theta}_{0},z_{i}\right)+\left(\hat{\boldsymbol{\theta}}-\boldsymbol{\theta}_{0}\right)^{\top}\cdot\left[\left.\frac{d\log p\left(\boldsymbol{\theta},z_{i}\right)}{d\boldsymbol{\theta}}\right|_{\boldsymbol{\theta}=\boldsymbol{\theta}_{0}}\right]+O\left(\hat{\boldsymbol{\theta}}-\boldsymbol{\theta}_{0}\right)^{2}.\label{eq:logp-series}
\end{gather}
By defining the vectors $\boldsymbol{\delta}\triangleq\hat{\boldsymbol{\theta}}-\boldsymbol{\theta}_{0}$
and $\kappa_{i}\triangleq\left.\frac{d\log p\left(\boldsymbol{\theta},z_{i}\right)}{d\boldsymbol{\theta}}\right|_{\boldsymbol{\theta}=\boldsymbol{\theta}_{0}}$,
we can rewrite (\ref{eq:log-link}) as
\begin{gather}
\log\mathbb{E}\left[n_{i}|\boldsymbol{\beta},\boldsymbol{\lambda},\hat{p}_{i}\right]=\log a_{i}\hat{p}_{i}+X_{i}\boldsymbol{\beta}+\kappa_{i}\boldsymbol{\delta}+O\left(\boldsymbol{\delta}^{2}\right).\label{eq:link-with-delta}
\end{gather}
In Supplementary Materials C, we show that the approximations in
our approach do not affect the asymptotic order of accuracy. Specifically,
the Laplace approximation that underlies REML estimation is accurate
to $O\left(n^{-1}\right)$ and our approximations are of the same
order (see Supplementary Materials C for the meaning of $n$). 

We have approximately that $\boldsymbol{\theta}_{0}\vert\mathbf{y}\sim N\left(\hat{\boldsymbol{\theta}},\mathbf{V}_{\boldsymbol{\theta}}\right)\implies\boldsymbol{\delta}\sim N\left(\mathbf{0},\mathbf{V}_{\boldsymbol{\theta}}\right),$
where the covariance matrix $\mathbf{V}_{\boldsymbol{\theta}}$ is
calculated as the negative inverse Hessian of (\ref{eq:detfct-ll}).
In other words, the ``posterior distribution'' of $\boldsymbol{\theta}$
from fitting the detection function now becomes a prior distribution
for $\rho$. To first order, $\boldsymbol{\delta}$ then plays the
same structural role in (\ref{eq:link-with-delta}) as the basis coefficients
$\boldsymbol{\beta}$. The design matrix for $\boldsymbol{\delta}$
($\kappa$ in (\ref{eq:link-with-delta})) is obtained by differentiating
the log-detection probabilities, with respect to $\boldsymbol{\theta}$
at $\hat{\boldsymbol{\theta}}$. Simple 3-point numerical differentiation
is perfectly adequate for calculation of the derivatives. $\mathbf{V}_{\boldsymbol{\theta}}$
should be readily available from detection function fitting (via the
Hessian) regardless of the complexity of the model. 

This method can be applied automatically to almost any distance sampling
setup provided one can calculate detection probabilities, find their
derivatives, and obtain a Hessian for the likelihood. Simultaneous
estimates $\boldsymbol{\beta}$ and $\boldsymbol{\delta}$ can be
obtained from standard GAM fitting software. Posterior inferences
about $\boldsymbol{\beta}$ (therefore $\rho$ and abundance) automatically
propagate the uncertainty from fitting the detection function.

The only technical difference from fitting a standard GAM, is that
$\boldsymbol{\lambda}$ is usually unknown and has to be estimated
(i.e., the prior on $\boldsymbol{\beta}$ has known covariance, but
unknown scale), whereas the prior on $\boldsymbol{\delta}$ is completely
determined from the detection function fitting (i.e., in effect $\lambda_{\boldsymbol{\delta}}=1/\phi$,
where $\phi$ is the scale parameter). This setup cannot be specified
directly in the \textsf{R} package \texttt{mgcv} because of implementation
details (at least up to version 1.8; it may be possible within other
GAM implementations), unless $\phi$ is fixed rather than estimated.
This is fine for Poisson or negative binomial response, but in our
experience, better fits can often be obtained using a Tweedie response
distribution, for which $\phi$ must be estimated. In order to implement
(\ref{eq:link-with-delta}) for a general response distribution using
\texttt{mgcv}, we therefore use a one-dimensional search over $\phi$
to maximize the marginal REML. At each iteration, given the working
value $\phi^{*}$, we re-fit the GAM fixing $\phi=\phi^{*}$ and $\lambda_{\boldsymbol{\delta}}=1/\phi^{*}$.
Speed can be improved by re-using some of the setup computations (design
matrices, etc) at each iteration.

\textbf{Diagnostics}\textit{\emph{. If the detection function fits
properly and the spatial model has adequate flexibility, then the
second-stage model should not lead to much change in the detection
function parameters, so that $\hat{\boldsymbol{\delta}}$ should be
``close'' to 0. Nevertheless, there is scope for interaction if
the detection function includes covariates that also vary systematically
over space. For example, if weather is systematically worse in some
parts of the survey region, then both $\boldsymbol{\beta}$ and $\boldsymbol{\theta}$
will contribute to the expected pattern of sightings, and the two
sets of parameters are partially confounded. (That is of course also
true for all-in-one models, as well for our two-stage model.)}}

\textit{\emph{There are several diagnostics that we have found useful
for checking consistency between the two parts of the model.}} The
first is to compare the inferred spatial distribution and abundance
from fitting (\ref{eq:log-link}) with the ``na{\"i}ve'' estimates
where detection uncertainty is ignored and the offset $a_{i}\hat{p}_{i}$
is treated is exact, ensuring that there are not large differences
in the estimated spatial distribution. The second is to check whether
the detection probabilities (by covariate level) would be substantially
changed by fitting the spatial model; in other words, whether $\hat{\boldsymbol{\delta}}$
is close enough to zero given its prior distribution, or, perhaps
more usefully, whether the overall detectability by covariate level
has changed. Since the fitted spatial model still includes the information
from the first stage, any shift of more than about 1 standard deviations
(based on the covariance from the detection function stage) might
merit investigation. Third, as a general diagnostic tool for density
surface models, we have found it useful to compare total observed
and expected numbers of sightings, grouped by detection covariates
(e.g., Beaufort). This can be helpful in diagnosing detection function
problems, e.g., failure of assumed certain detectability at zero distance
under poor weather conditions, as well as failures of the spatial
model (e.g., an abrupt change in density). In addition one could also
use standard detection function model checking (e.g., quantile-quantile
plots) with the adjusted parameters, $\hat{\boldsymbol{\theta}}+\hat{\boldsymbol{\delta}}$.

\textbf{Calculating $\text{Var}(\hat{N})$}. Once detection function
uncertainty has been propagated, we only need to deal with uncertainty
in the GAM, which now has an updated covariance matrix. We therefore
can rely on two commonly-used methods to obtain the variance of model
outputs like abundance, $\hat{N}$.

\textbf{1. Delta method: }We can calculate: 
\begin{equation}
\text{Var}(\hat{N})=\left(\mathbf{a}_{p}\frac{\partial\exp\mathbf{X}_{p}\boldsymbol{\beta}}{\partial\boldsymbol{\beta}}\Bigg\vert_{\boldsymbol{\beta}=\boldsymbol{\hat{\beta}}}\right)\boldsymbol{V}_{\hat{\bm{\beta}}}\left(\mathbf{a}_{p}\frac{\partial\exp\mathbf{X}_{p}\boldsymbol{\beta}}{\partial\boldsymbol{\beta}}\Bigg\vert_{\boldsymbol{\beta}=\boldsymbol{\hat{\beta}}}\right)^{\text{\ensuremath{\intercal}}},\label{eq:delta}
\end{equation}
(the delta method) where $\mathbf{V_{\hat{\bm{\beta}}}}$ is the
covariance matrix for the GAM coefficients \citet[Sections 5.8 \& 6.9.3]{wood2017generalized}.
We form the prediction matrix, $\mathbf{X}_{p}$, which maps model
coefficients to values of the linear predictor for the prediction
data, so $\boldsymbol{\hat{\eta}}_{p}=\mathbf{X}_{p}\boldsymbol{\hat{\beta}}$
\citep[Section 6.10]{wood2017generalized}. Derivatives are evaluated
at the estimated values of the model parameters.

\textbf{2. Posterior simulation: }The posterior for $\boldsymbol{\beta}$
given data $\mathbf{y}$ and smoothing parameters $\bm{\lambda}$,
are approximately distributed as $\boldsymbol{\beta}\vert\mathbf{y},\boldsymbol{\lambda}\sim N(\hat{\boldsymbol{\beta}},\mathbf{V_{\hat{\bm{\beta}}}})$.
The following algorithm then can be used:
\begin{enumerate}
\item For $b=1,\ldots,B$:
\begin{enumerate}
\item Simulate from $N(\hat{\boldsymbol{\beta}},\mathbf{V}_{\hat{\bm{\beta}}})$,
to obtain $\hat{\boldsymbol{\beta}_{b}}$.
\item Calculate predicted abundance, $\hat{N}_{b}=\mathbf{a}_{p}\exp(\mathbf{X}_{p}\hat{\boldsymbol{\beta}_{b}})$
(where $\mathbf{a}_{p}$ is a row vector of areas for the prediction
cells).
\end{enumerate}
\item Calculate the empirical variance or percentiles of the $\hat{N}_{b}$s.
\end{enumerate}
In practice $B$ in the order of 1000s appears to work well, though
there may be some issues when the approximation breaks down. In these
cases we recommend the use of importance sampling (either using importance
weights to calculate weighted summaries or using a second resampling
of the $\hat{N}_{b}$s) or a Metropolis-Hastings sampler (as implemented
in \texttt{mgcv::gam.mh}). Further examples are given in Supplementary
Materials.

\textbf{Software.} The procedure given in this section is implemented
in the R package \texttt{dsm}, available on CRAN. The \texttt{dsm\_varprop}
function in the package allows the user to provide a fitted DSM and
a prediction grid. Using the delta method it will then calculate an
uncertainty estimate for the estimated abundance for that prediction
grid. The function also returns the refitted GAM so one can extract
the full covariance matrix and perform posterior simulation if required.
Diagnostics for $\hat{\boldsymbol{\delta}}$ are calculated by a \texttt{summary}
method for the returned object.

\section{Previous methods for estimating uncertainty in Density Surface Models}

\label{sec:prev}Several approaches have previously been suggested
to combine detection function and spatial model predicted abundance
uncertainties; we review them briefly here. We need to estimate the
following:
\begin{align*}
\mathrm{Var_{P}}(\text{log}N) & =\mathbb{E}_{P}[\mathrm{Var}(\text{log}N\vert P)]+\mathrm{Var}_{P}[\mathbb{E}(\text{log}N\vert P)]\\
 & \approx\text{Var}(\text{log}N\vert\{\hat{p}_{i};i=1,\ldots,n\})+\mathrm{Var}_{P}[\text{log}\hat{N}(\{\hat{p}_{i};i=1,\ldots,n\})],
\end{align*}
where $P$ here is a random variable for the (uncertain) probability
of detection and the subscripts indicate the expectation/variance
taken over that variable. $\hat{N}(\{\hat{p}_{i};i=1,\ldots,n\})$
is the estimated abundance as a function of estimated detection probabilities.
The first part of this can be derived from GAM theory as shown in
the previous section; the second is more tricky. 

\textbf{Assuming independence. }When $\hat{p}_{i}$ is the same for
all observations, then $N(\hat{p})\propto1/\hat{p}$, so $\hat{N}$
and $\hat{p}$ are independent. The total variance of the abundance
estimate can be calculated by combining the GAM variance estimate
with the variance of the probability of detection summing the squared
coefficients of variation ($\text{CV}(X)=\sqrt{\text{Var}(X)}/\bar{X}$)
\citep{Goodman:1960dd}. Hence 
\[
\text{Var}_{\text{IND}}(\hat{N})=\frac{\hat{N}^{2}}{\text{CV}^{2}(\hat{N}_{\text{GAM}})+\text{CV}^{2}(\hat{p})}.
\]
When there are not covariates in the detection function we calculate:
\begin{equation}
\text{Var}(\hat{p})=\left(\frac{\partial\hat{p}}{\partial\boldsymbol{\theta}}\Bigg\vert_{\boldsymbol{\theta}=\boldsymbol{\hat{\theta}}}\right)\boldsymbol{V}_{\boldsymbol{\hat{\theta}}}\left(\frac{\partial\hat{p}}{\partial\boldsymbol{\theta}}\Bigg\vert_{\boldsymbol{\theta}=\boldsymbol{\hat{\theta}}}\right)^{\text{T}}.\label{eqn-deltamethod}
\end{equation}
This is fine when the detection function does not contain any covariates,
as there is no covariance then between the effort and density covariates
(the procedure outlined in Section \ref{sec:varprop} does not yield
a different answer). In the case where detectability is a function
of covariates it is impossible in general to justify the use of the
CV decomposition as there are correlations between the spatial distribution
and the covariates affect detectability. 

The approach taken by Program Distance \citep{Thomas:2010cf} is to
use \foreignlanguage{british}{Horvitz-Thompson-adjusted} counts per
segment, instead of the observed count, as the response in the GAM.
Thus removes the detectability from the right hand side of (\ref{dsm-eqn}).
Variance is then calculated by taking the probability of detection
averaged over the observations by first calculating the Horvitz-Thompson
estimate of the abundance in the covered area ($\hat{N}=\sum_{i}g_{i}/\hat{p_{i}}$,
where $g_{i}$ is group size of the $i^{\text{th}}$ observation and
$\hat{p_{i}}$ is the probability of detecting that group) then using
that $\hat{N}$ to calculate the implied average detectability, had
the analysis not contained covariates ($\tilde{p}=\tilde{n}/\hat{N}$,
where $\tilde{n}$ is the number of observed groups). The numerical
derivatives of $\tilde{p}$ with respect to $\boldsymbol{\theta}$
can then be used in (\ref{eqn-deltamethod}) to derive a variance
for this probability of detection, averaged over the observations.

We do not recommend this approach either. Transforming the response
through multiplication by a random variable breaks the mean-variance
and independence assumptions of the GAM, so that the computed $\text{CV}\left(\hat{N}_{\text{GAM}}\right)$
is invalid when detection covariates are present. Additionally, there
is no coherent way to generalize the formula to small-area predictions
--- the effort covariates within a small area will not have the same
range as those in the larger survey area (e.g., weather conditions
will not be homogenous throughout the survey area). Hence, the uncertainty
that applies to the overall $\tilde{p}$ is usually not the appropriate
uncertainty to apply to a small area where observing conditions may
be atypical.

\textbf{The bootstrap. }Bootstraps are sometimes seen as an attractive
alternative to deal with all aspects of variance in DSMs. \citet{Hedley:2004et}
describe two possible implementations (one parametric, one non-parametric),
which are not easy to choose between and which do not necessarily
give similar answers. Ignoring computational time issues, the first
practical difficulty in setting up a ``good'' non-parametric bootstrap
for a DSM is sampling units, independent of the fitted model.

The second, more substantial, issue is the fundamental statistical
problem with combining smoothers with bootstraps. The problem does
not seem to be well-known in the statistical ecology literature, so
we give an explanation here. The basic problem is that (most) bootstraps
use only the posterior modes of random effects (smooths), thus omitting
a key part of the posterior uncertainty. To see this, consider a simple
``spatial model'' where the region is divided into blocks, each
with its own independent random effect, and a bootstrap that generates
new data at each original observation/transect, either parametrically
or non-parametrically. If one of the blocks is unsampled in the original
data, it will be unsampled in every realization too, and the ``spatial
model'' simply sets the point estimate of that random effect to zero
in \emph{every} bootstrap realization; hence a bootstrap will ascribe
zero uncertainty for the density in that block. The correct inference
would of course be for the random effect to retain its prior variance. 

\begin{figure}
\begin{knitrout}
\definecolor{shadecolor}{rgb}{0.969, 0.969, 0.969}\color{fgcolor}
\includegraphics[width=\maxwidth]{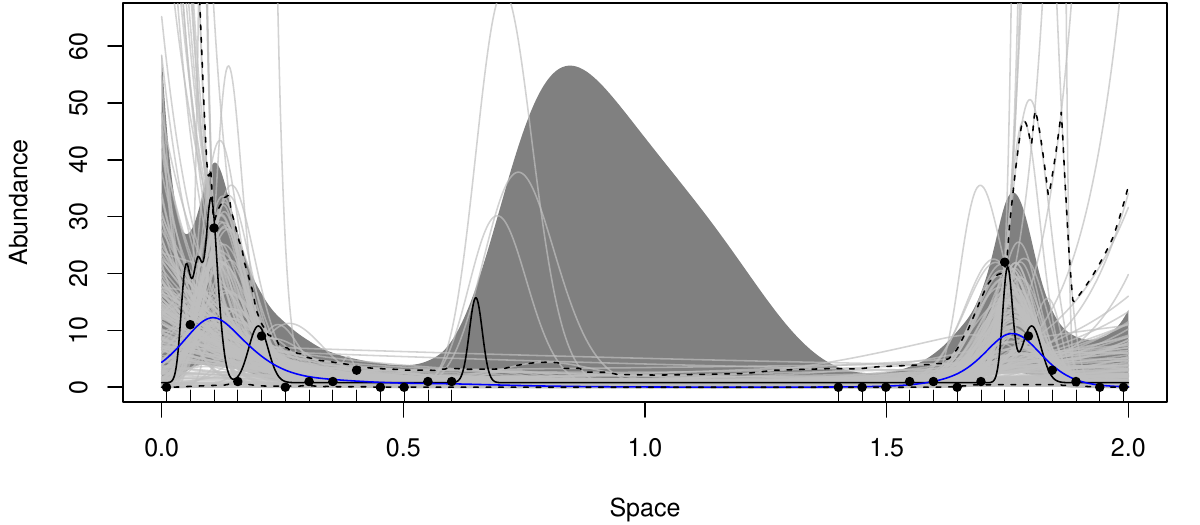} 

\end{knitrout}

\caption{Comparison of bootstrap and analytical uncertainty for a Poisson process.
The black line is the true intensity function (on the response scale)
and points are observations. Blue line is a smooth of space and light
grey wiggly lines are 500 bootstrap predictions, dashed lines are
point-wise upper and lower 95\% quantiles from the bootstrap, the
dark grey band is the analytical GAM confidence band using (\ref{eq:delta}).
The bootstrap appears confident that there is nothing in the unsampled
area, but the analytical estimate illustrates how little we know.\label{fig:bootin}}
\end{figure}

This phenomenon has been well-known in statistics since at least \citet{Laird:1987jw}
(see also the discussants), who coined the term ``na{\"i}ve bootstrap''
for such procedures that ignore the point estimate shrinkage inevitable
in mixed or random effect models (fixed effect models are not susceptible
in the same way). They proposed some parametric modifications (``type
II'' and ``type III'' bootstraps) that are more effective in the
\textit{\emph{IID}} and block-structured situations that they consider.
However, the underlying theory is complex \citep{Carlin:1991hz,Carlin:2008wa}
and it is far from clear whether simple yet reliable bootstraps can
be devised for complicated multi-stage random effect situations like
DSMs. Figure \ref{fig:bootin} shows a simple unidimensional Poisson
process, sampled at either end but not in the middle (rug plot). Bootstrap
replicates (shown in light grey, of which there are 500) largely fail
to capture our uncertainty in the unsampled middle area. The analytical
estimate (dark grey band) illustrates how little we know about the
unsampled area.

The above does not imply that simple or indeed complicated bootstraps
will \emph{never} give reliable results in DSMs; given plenty of observations
and good, uniform coverage, many approaches to inference will give
similar and good results. However, it is sometimes not obvious whether
this holds for a specific dataset, nor what to do bootstrap-wise if
not. Instead, the (empirical) Bayesian framework of GAMs offers a
coherent and general-purpose way to capture uncertainty.

\subsubsection*{}

\section{A new model for group size}

\label{sec:group-size-model}Our variance propagation method so far
works if detectability depends only on effort covariates, but not
for individual covariates such as group size. Incorporating individual
covariates in the detection function is not problematic but it is
not obvious how to allow for these different detection probabilities
in the GAM. Further, it is not obvious how to combine predictions
of different group sizes since average group sizes may vary spatially.

One approach is to use the Horvitz-Thompson-adjusted response described
in the previous section, but as mentioned above this does not allow
variance propagation. One could fit separate spatial models to subsets
of the data for each group size, but it seems inefficient to not share
information between subsets of the data. Next, we show instead how
to extend our variance-propagation method to deal with group size.

We form $M$ categories of group sizes, denoted $\{g_{m};m=1,\ldots,M\}$,
where groups within each category have similar detectability, and
fit a detection function incorporating these group size categories.
We then fit a GAM to an $M$-fold replicate of the dataset, with the
response in the $m^{\text{th}}$ replicate of the $i^{\text{th}}$
segment being $n_{im}$, the number of groups in category $m$ that
were seen in that segment. (The total number of observations is unchanged;
each observation is allocated to just one of the ``replicates''.)
Group size category (as a factor) is included as an explanatory variable,
and smooths are modified to allow similar variations in density of
groups with different sizes. There are no extra assumptions in this
formulation from the model in Section \ref{sec:varprop}, except to
assume that the numbers of groups of different size categories in
a given segment are independent, given the underlying density (which
is allowed to vary with group size).

\textbf{Factor-smooth interactions.} We extend (\ref{dsm-eqn}) to
include multiple smooths of space which correspond to different categorizations
of group size, so our model is:
\begin{equation}
\mathbb{E}\left[n_{i,g_{m}}|\boldsymbol{\beta},\boldsymbol{\lambda},p(\hat{\boldsymbol{\theta}};z_{i},g_{m})\right]=a_{i}p(\hat{\boldsymbol{\theta}};z_{i},g_{m})\exp\left(\beta_{0}+f_{x_{1},g_{m}}(x_{i1})+\sum_{k=2}^{K}f_{k}(x_{ik})\right),\label{eq:fs-DSM}
\end{equation}
for $m=1,\ldots,M$ where $n_{i,g_{m}}$ is the number of observed
groups in group class $g_{m}$ in segment $i$ and $f_{x_{1},g_{m}}$
is the spatial smooth (where $x_{1}$ is a spatial coordinate) for
group size class $g_{m}$. Smoothers like $f_{x_{1},g_{m}}$ are referred
to as \textit{factor-smooth interactions} \citep{wood2017generalized,10.7287/peerj.preprints.27320v1}.
$f_{k}$ are any other smooths (of covariates $x_{k}$, for $k>1$).
For clarity we make the dependence on group size class explicit: $p(\hat{\theta};z_{i},g_{m})$,
i.e., the probability of detection given segment-level detection covariates
$z_{i}$ and group size $g_{m}$. 

There are a number of different possible forms for $f_{x_{1},g_{m}}$.
These vary in two main ways: (1) do levels share a smoothing parameter,
or have separate ones? (2) do smooths tend toward a ``global'' smooth
that dictates a general spatial effect? Here we adopt the ``\texttt{fs}''
basis in \texttt{mgcv} which can be thought of as a smooth version
of a random slopes model: smooths are generated for each factor level
with smooths defined as deviations from a reference level, with all
smooths sharing the same smoothing parameter. This is appealing as
we might expect that the spatial smooths for each group size are similar
but there might be some process that generates larger groups in certain
places (e.g., large prey aggregations attracting large groups of animals).
This approach is easily extended to other density covariates (e.g.,
$x_{1}$ could be bathymetry or vegetation cover).

\textbf{Abundance and uncertainty estimation with group size smooths}.
Abundance is estimated by summing over the predictions for each group
size category ($\hat{N}_{m}$) and weighting them by the corresponding
mean group size ($\bar{g}_{m}$): $\hat{N}=\sum_{m=1}^{M}\bar{g}_{m}\hat{N}_{m}.$
We can find $\text{\text{Var}(}\hat{N}\vert\bar{G)}$ (where $\bar{G}$
is the mean group size) from the variance propagation procedure above,
but we need $\text{\text{Var}(}\hat{N})$, which we can obtain from
the Law of Total Variance:
\begin{align}
\mathrm{Var}(\hat{N}) & =\mathbb{E}_{\bar{G}}[\mathrm{Var}(\hat{N}\vert\bar{G})]+\mathrm{Var}[\mathbb{E}_{\bar{G}}(\hat{N}\vert\bar{G})]\nonumber \\
\text{} & =\mathrm{Var}(\hat{N}\vert\bar{G})+\sum_{m=1}^{M}\mathrm{Var}(\bar{G}_{m})\hat{N}_{m}^{2},\label{eq:varN-group}
\end{align}
where $\mathrm{Var}(\bar{G}_{m})$ reflects the uncertainty about
mean group size within a category, to be estimated empirically from
all the observed groups in that category. The effect of $\mathrm{Var}(\bar{G}_{m})$
on $\mathrm{Var}(N)$ should be small (because categories are narrow,
and mean must lie within category), and also should not vary much
spatially, so no further spatial adjustment to that variance component
is required. 

\section{Examples}

\label{examples}\textbf{Island Scrub-Jays}. We first apply our variance
propagation method in a simple situation where there is covariance
between the abundance and detection processes, that is the case of
a spatially-varying detection covariate. Island Scrub-Jays (\textit{Aphelocoma
insularis}) are endemic to Santa Cruz Island, California. Jays primarily
reside in areas of chaparral and forest, though the density of this
foliage also affects detectability. \citet{sillett_hierarchical_2012}
model the distribution Island Scrub-Jays from 307 point transects
surveyed in fall 2008 and spring 2009. Distances were binned into
three intervals due to responsive movement ($[0m-100m],(100m-200m],(200m-300m]$).
Proportion chaparral ($\texttt{chap}$) and proportion forest ($\texttt{forest}$)
were available as covariates, as was elevation ($\texttt{elev}$).
\citeauthor{sillett_hierarchical_2012} fitted a hierarchical model
assuming a negative binomial distribution for abundance and a multinomial
detection process using a half-normal detection function. Their best
models (by AIC) were: fall 2008 abundance modelled as $\beta_{0}+\beta_{1}\texttt{\texttt{chap}}^{2}+\beta_{2}\texttt{chap}+\beta_{3}\texttt{elev}$,
with detectability as a function of $\texttt{chap}$; spring 2009
abundance modelled as $\beta_{0}+\beta_{1}\texttt{chap}^{2}+\beta_{2}\texttt{chap}+\beta_{3}\texttt{elev}^{2}+\beta_{4}\texttt{elev}$,
detectability as a function of $\texttt{forest}$. 

We replicated the analysis of \citeauthor{sillett_hierarchical_2012}
using our two-stage variance propagation approach to show that our
method can be used in such a situation. In summary, final coefficient
estimates were very close to those in the original paper, abundance
estimates with associated 95\% CIs were very similar for both seasons:
Fall 2008 DSM $\hat{N}=$2272 (1625--3175), Sillet et al. $\hat{N}=$2267
(1613--3007) and Spring 2009 DSM $\hat{N}=$1684 (1263-2246), Sillet
et al. $\hat{N}=$1705 (1212--2369); Supplementary Material A gives
the comparison in full. For the spring model, the value of the $\texttt{forest}$
coefficient in the detection function changed effect size from -0.18
(SE=0.06) to -0.083 (SE=0.062) after propagation (indicating no issue
with our $\hat{\boldsymbol{\delta}}$ diagnostic). \textit{\emph{By
giving the GAM the flexibility to slightly adjust the detection function
parameters via }}$\hat{\boldsymbol{\delta}}$\textit{\emph{ (as opposed
to treating the estimated detection probabilities as certain), the
CV of the abundance estimate is actually improved in this case, from
18.4\% to 14.8\%. }}

The jay data presents a particularly interesting case as the covariates
in the GAM are fixed effects, there is therefore no ``cost'' (in
terms of the penalized likelihood) to changing the GAM coefficients.
We see minimal changes in the parameters of the fall model (Supplementary
Material A, Table 1), as these are already well modelled (no doubt
due to the good coverage of the data): the detection function includes
$\texttt{chap}$ and the GAM includes $\texttt{chap}$ and $\texttt{chap}^{2}$,
so any adjustment via $\hat{\boldsymbol{\delta}}$ is a 3$^{\text{rd}}$
order effect. The spring model has different covariates in each model
component, making the correction necessary.

The survey design had extremely good coverage over Santa Cruz Island.
We decided to see what the effect of ``unbalancing'' the design
would be to test the robustness of our model. We randomly subsampled
the fall data to contain only 100 sites, then removed those where
chaparral cover was greater than the mean chaparral proportion (over
all points). Our subsample was left with 16 detections at 65 points.
Fitting the fall DSM to the reduced data yields $\hat{N}=$26,434
(95\% CI 209-3,349,000; CV=2,100\%). Post-variance propagation, we
obtain $\hat{N}=$2,831 (95\% CI 39-206,000; CV=1,100\%), both detection
function and GAM coefficients having changed (see Supplementary Material
A, Table 3). While we would expect a high variance for such a small
and unbalanced dataset (and indeed we obtain this), our procedure
tames the model to an extent, giving a more realistic estimate of
abundance. Once information about both model components is allowed
to inform the parameter estimation simultaneously, the coefficients
are corrected.

\textbf{Island Scrub-Jays Simulation}. To assess performance of our
variance propagation method with the delta method and a one stage
fully Bayesian approach we conducted a simulation using the Island
Scrub-Jay data as a starting point. We kept spatial coverage constant
throughout the simulation settings but varied the detectability and
therefore the number of observations available for the detection function
component of the model. Full details of the simulation setup are given
in Supplementary Material B. Here we note that our variance propagation
method performed well in terms of bias in the abundance estimate and
its corresponding variance estimate compared to the fully Bayesian
model, even when sample size decreased (Supplementary Material B,
Figure 2).

\begin{figure}
\begin{knitrout}
\definecolor{shadecolor}{rgb}{0.969, 0.969, 0.969}\color{fgcolor}
\includegraphics[width=\maxwidth]{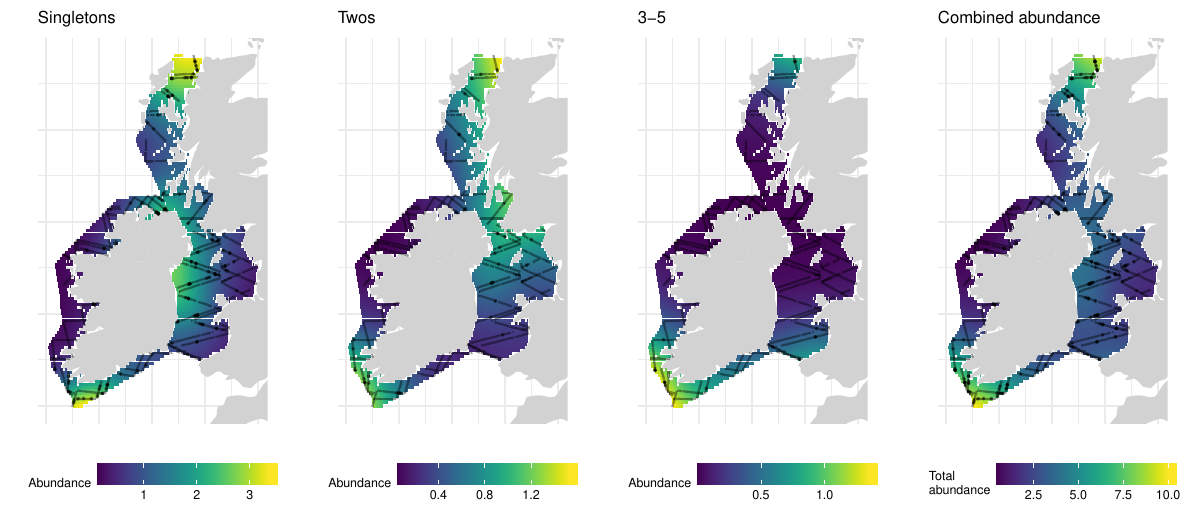} 

\end{knitrout}

\caption{Predicted density surfaces from the new group size model for harbour
porpoise. First three plots are density maps for the given group size
(i.e., group abundance multiplied by mean group size), right plot
shows the combined map, summing the previous three plots per prediction
cell. We can see that distribution is roughly similar in all three
group size categories though with almost no larger groups in the North,
far more animals occurring as singletons than in larger groups. \label{fig:hp-pred}}
\end{figure}

\textbf{Harbour porpoise.} To illustrate our new group size model,
we re-analyse an aerial line transect survey of harbour porpoise in
Irish Sea, coastal Irish waters and Western coastal Scotland, where
we see spatial variation in observed group size of 1 to 5 animals
(typical for harbour porpoise; \citealp[e.g.,][]{siebert2006decade};
points in Figure \ref{fig:hp-pred}). During SCANS-II aerial surveys,
two observers recorded cetacean detections (along with sighting conditions)
from bubble windows on both sides of a plane flying at 183m. Complete
survey details and a comprehensive analysis is given in \citet{Hammond:2013fs}.
For simplicity we assume certain detection on the trackline, no errors
in group size estimation (less likely with aerial than in shipboard
surveys for harbour porpoise; Phil Hammond, Debi Palka, \textit{pers.
comm.}, November 2017), and negligible island/coastline effects in
the spatial model. 

To fit our DSM, three group size bins were formed: size 1 (131 observations),
2 (35 observations) and 3-5 (14 observations). A hazard-rate detection
function was fitted to the observed distances (truncated at 300m)
with the group size bin ($g_{m}$, $m=1,\ldots,3$) and Beaufort ($B_{i}$,
binned as 0-1, 2 and 3-5) as factor covariates. Detectability for
each segment $i$ per group size factor was then estimated from the
detection function: $p(\hat{\boldsymbol{\theta}};B_{i},g_{m})$. Following
(\ref{eq:fs-DSM}) we fitted the DSM: 
\[
\mathbb{E}\left[n_{i,g_{m}}|\boldsymbol{\beta},\boldsymbol{\lambda},p(\hat{\boldsymbol{\theta}};B_{i},g_{m})\right]=a_{i}p(\hat{\boldsymbol{\theta}};B_{i},g_{m})\exp\left(\beta_{0}+f_{E,N,g_{m}}(E_{i},N_{i})\right),
\]
where in segment $i$ of area $a_{i}$ the observed number of groups
in size category $g_{m}$ was denoted $n_{i,g_{m}}$. It was assumed
the response was Tweedie-distributed where the power parameter was
constrained to be greater than $1.2$ to avoid numerical issues. Each
$f_{E,N,g_{m}}$ was a smooth of space (projected Easting/Northing;
$E_{i}$, $N_{i}$) for group category $g_{m}$ and had a maximum
basis size of 20 (total maximum basis size for $f_{E,N,g_{m}}$ was
therefore 60).

The fitted model had a total effective degrees of freedom of 20.47
for $f_{E,N,g_{m}}$. GAM checking showed reasonable fit to the data.
Table \ref{tab:hp-obs-exp} shows observed vs expected counts by Beaufort
--- there is some misfit at the highest state, perhaps because detection
probability at zero distance changes with Beaufort level (from \citeauthor{Hammond:2013fs}:
0.45 for Beaufort 0-1 and 0.31 in Beaufort 2-3). Plots of the per-group
size bin predictions and the combined prediction are given in Figure
\ref{fig:hp-pred}, which show some consistent patterns between group
size classes (``hotspot'' off Southern Ireland) and some differences
(varying distribution in the Irish Sea and Western Scotland), this
kind of insight is not possible using a single smooth for all observations
and may prove useful in cases where there are occasional very large
group sizes (e.g., oceanic dolphins). Using (\ref{eqn-deltamethod}),
the CV of abundance was estimated to be 2.36\%,
when our new variance propagation method was used the CV was estimated
as 9.65\%. The assumption of independence
(via (\ref{eqn-deltamethod})) underestimates uncertainty in the case
where group size (and detectability) vary in space. Only a small piece
of code implementing (\ref{eq:varN-group}) was required in addition
to the \texttt{dsm\_varprop} function (included in Supplementary Material
A), we then gain the ability to make inferences about group size spatial
distribution (traditionally requiring two separate models, one for
encounter rate, one for group size; e.g., \citealp{Becker:2014bl}),
as well as improve uncertainty estimation via variance propagation.

\begin{table}
\caption{Observed versus expected counts from the harbour porpoise DSM (post-variance
propagation) at levels of Beaufort used in the detection function.
\label{tab:hp-obs-exp}}
\begin{knitrout}
\definecolor{shadecolor}{rgb}{0.969, 0.969, 0.969}\color{fgcolor}
\begin{tabular}{c|rrrrr}

  & 0-1 & 2 & 3-5\\
\hline
Observed & 98.00 & 36.00 & 30.00\\

Expected & 96.59 & 35.13 & 35.69\\

\end{tabular}

\end{knitrout}
\end{table}

\section{Discussion}

\label{sec:Discussion}

\textit{\emph{Combining the uncertainty from detection functions with
that from spatial models has been a challenging problem for point
and line transect analysis, requiring either complicated bespoke software
that combines two model components, or ad hoc approaches that lack
statistical justification. In this paper, we have demonstrated a simple,
flexible, and statistically sound method that can (i) }}propagate
uncertainty from detectability models to the spatial models for a
particular class of detection function (i.e., those without individual-level
covariates)\textit{\emph{ and (ii)}}\emph{ }include group size as
a covariate in the detection function while still being able to propagate
uncertainty and address spatial variation in group size\textit{\emph{.
Our methods are implemented in}}\textit{ }the \texttt{dsm} package
for R but can be implemented in any standard GAM fitting software. 

It is straightforward to apply our factor-smooth approach group size
more generally to individual-level covariates which affect detectability
and vary in space, but do not directly affect \textit{\emph{abundance,
such as observable behaviour. For example, feeding groups might be
more (or less) conspicuous than resting groups, and the proportion
feeding\slash resting may vary across the surveyed region. Unless
detectability is included in the analysis, biased abundance estimates
could result, especially when survey coverage is non-uniform; and
there has been no simple way until now to include such effects in
the spatial model.}} A major advantage of our approach over simple
(or complex) stratification schemes is that we are now sharing information
between the levels of our categorized variable. This makes the results
less sensitive to over-specifying the number of categories, as the
model will shrink back towards the simpler model in the absence of
strongly informative data. We also note that the factor-smooth approach
could be applied to all smooth terms in the GAM, allowing for a very
flexible model. This would be appropriate only if it was reasonable
that all smooths vary according to the detectability covariate (e.g.,
feeding behaviour in our harbour porpoise example might depend on
both space and depth).

We have assumed that all variables are measured without much error.
Measurement error for individual-level covariates such as group size
can be a serious problem in distance sampling \citep{Hodgson:2017kk}---distance
between observer and group can affect not just detectability, but
also the extent of group size error. If group size varies spatially,
it is hard to see how to separate the spatial modelling stage from
the distance sampling stage. A full discussion is beyond the scope
of this paper, but we suspect that specially-designed observation
protocols and bespoke analyses may be the only way to tackle such
thorny cases.

All-in-one fitting of both detection and spatial models is also possible
\citep[e.g.,][]{Johnson:2009gf,sillett_hierarchical_2012,Yuan:2017}.
If models are specified correctly, then the all-in-one approach could
in theory be slightly more efficient, but only insofar as it takes
account of third-order changes in the detection function likelihood
(since our approach uses a quadratic approximation). That seems unlikely
to make much difference in general---and as is the case for the Island
Scrub-jay example. Our own preference is therefore to use the two-stage
approach, mainly because in our experience the careful fitting of
detection functions is a complicated business which can require substantial
model exploration and as few as possible ``distractions'' (such
as simultaneously worrying about the spatial model). The two-stage
process allows any form of detection function to be used, without
having to make deep modifications to software. In summary, if one
\textit{knew }one had the correct model to begin with, one-stage fitting
would be slightly more efficient, but this is never the case in practice.

It is valuable to check for any tension or confounding between the
detection function and density surface parts of the model, which can
occur if there are large-scale variations in sighting conditions across
the survey region, and which is readily diagnosed in a two-stage model.
Although this does not appear to lead to problems in the datasets
we have analysed with the software described in this paper, we have
come across it in other variants of line transect-based spatial models
with different datasets.  It may not be so easy to detect partial
confounding when using all-in-one frameworks.

Finally, we note that the approach outlined here (of using a first-stage
estimate as a prior for a second-estimate, and propagating variance
appropriately) is quite general and is comparable to standard sequential
Bayesian approaches to so-called ``integrated data models''. The
first-stage model need not be a detection function but instead could
be from another GAM (or other latent Gaussian model). Again, this
allows us to ensure that first-stage models are correct before moving
to more complex modelling. Modelling need not only be two-stage and
could extend to multi-stage models \citep{hooten2018prior}.

\section*{Acknowledgements}

The authors thank Natalie Kelly, Jason Roberts, Eric Rexstad, Phil
Hammond, Steve Buckland and Len Thomas for useful discussions, Devin
Johnson for the suggestion of this as a general statistical method,
and Simon Wood for continued development of \texttt{mgcv} and GAM
theory. The manuscript was great improved by comments from the editor
and two anonymous reviewers. Data from the SCANS-II project was supported
by the EU LIFE Nature programme (project LIFE04NAT/GB/000245) and
governments of range states: Belgium, Denmark, France, Germany, Ireland,
Netherlands, Norway, Poland, Portugal, Spain, Sweden and UK. This
work was funded by OPNAV N45 and the SURTASS LFA Settlement Agreement,
and being managed by the U.S. Navy's Living Marine Resources program
under Contract No. N39430-17-C-1982, US Navy, Chief of Naval Operations
(Code N45), grant number N00244-10-1-0057 and the International Whaling
Commission.

\bibliographystyle{abbrvnat}
\bibliography{full.bib}

\end{document}


\title{Supplementary Material A: Island Scrub-jay analysis}
\author{Mark V. Bravington\\
David L. Miller\\
Sharon L. Hedley}
\maketitle

\section{Introduction}

This analysis is based on data from the paper:

\textit{Sillett, T. Scott, Richard B. Chandler, J. Andrew Royle, Marc
Kery, and Scott A. Morrison. \textquoteleft Hierarchical Distance-Sampling
Models to Estimate Population Size and Habitat-Specific Abundance
of an Island Endemic\textquoteright . Ecological Applications 22,
no. 7 (2012): 1997-2006}

Data is available at: \url{https://figshare.com/articles/Supplement_1_R_code_data_and_grid_covariates_used_in_the_analyses_/3517754 }.
Script \texttt{get\_issj\_data.R} converts this to a \texttt{dsm}-compatible
format (or use \texttt{issj.RData}).

The authors analyse 2 point transect surveys of Island Scrub-jays
from Santa Cruz Island, CA. Each survey consists of 307 sample locations
(point transects), one ocurring in fall 2008 and the other in spring
2009. Distances were binned into 3 intervals. Locations of the points
were available, along with vegetation (proportion forest and proportion
chaparral) and elevation. They dredged the data (see their table 2),
but their two final models were:
\begin{description}
\item [{Fall~2008~model:}] abundance was given as $\beta_{0}+\beta_{1}\texttt{chaparrel}^{2}+\beta_{2}\texttt{chaparrel}+\beta_{3}\texttt{elevation}$,
detectability as a function of $\texttt{chaparral}$.
\item [{Spring~2009~model:}] abundance was given as $\beta_{0}+\beta_{1}\texttt{chaparrel}^{2}+\beta_{2}\texttt{chaparrel}+\beta_{3}\texttt{elevation}^{2}+\beta_{4}\texttt{elevation}$,
detectability as a function of $\texttt{forest}$.
\end{description}
In both cases, abundance was assumed negative binomially distributed
and a half-normal detection function was fitted.

\section{Analysis}

Reproducing the analysis from Sillet et al. First loading the requisite
data and R packages.

\begin{knitrout}
\definecolor{shadecolor}{rgb}{0.969, 0.969, 0.969}\color{fgcolor}\begin{kframe}
\begin{alltt}
\hlcom{# load data}
\hlkwd{load}\hlstd{(}\hlstr{"issj.RData"}\hlstd{)}

\hlcom{# load packages}
\hlkwd{library}\hlstd{(Distance)}
\end{alltt}

{\ttfamily\noindent\itshape\color{messagecolor}{\#\# Loading required package: mrds}}

{\ttfamily\noindent\itshape\color{messagecolor}{\#\# This is mrds 2.2.4\\\#\# Built: R 4.0.2; ; 2020-11-30 17:31:53 UTC; unix}}

{\ttfamily\noindent\itshape\color{messagecolor}{\#\# \\\#\# Attaching package: 'Distance'}}

{\ttfamily\noindent\itshape\color{messagecolor}{\#\# The following object is masked from 'package:mrds':\\\#\# \\\#\#\ \ \ \  create.bins}}\begin{alltt}
\hlkwd{library}\hlstd{(dsm)}
\end{alltt}

{\ttfamily\noindent\itshape\color{messagecolor}{\#\# Loading required package: mgcv}}

{\ttfamily\noindent\itshape\color{messagecolor}{\#\# Loading required package: nlme}}

{\ttfamily\noindent\itshape\color{messagecolor}{\#\# This is mgcv 1.8-33. For overview type 'help("{}mgcv-package"{})'.}}

{\ttfamily\noindent\itshape\color{messagecolor}{\#\# Loading required package: numDeriv}}

{\ttfamily\noindent\itshape\color{messagecolor}{\#\# This is dsm 2.3.1.9007\\\#\# Built: R 4.0.2; ; 2020-10-30 08:32:41 UTC; unix}}\end{kframe}
\end{knitrout}

\subsection{Fall 2008}

Running models:

\begin{knitrout}
\definecolor{shadecolor}{rgb}{0.969, 0.969, 0.969}\color{fgcolor}\begin{kframe}
\begin{alltt}
\hlcom{# fit detection function}
\hlstd{df_fall} \hlkwb{<-} \hlkwd{ds}\hlstd{(obs_fall,} \hlkwc{transect}\hlstd{=}\hlstr{"point"}\hlstd{,} \hlkwc{formula}\hlstd{=}\hlopt{~}\hlstd{chaparral)}
\end{alltt}

{\ttfamily\noindent\itshape\color{messagecolor}{\#\# Columns "{}distbegin"{} and "{}distend"{} in data: performing a binned analysis...}}

{\ttfamily\noindent\itshape\color{messagecolor}{\#\# Model contains covariate term(s): no adjustment terms will be included.}}

{\ttfamily\noindent\itshape\color{messagecolor}{\#\# Fitting half-normal key function}}

{\ttfamily\noindent\itshape\color{messagecolor}{\#\# AIC= 330.658}}

{\ttfamily\noindent\itshape\color{messagecolor}{\#\# No survey area information supplied, only estimating detection function.}}\begin{alltt}
\hlcom{# fit spatial model}
\hlstd{ll_fall} \hlkwb{<-} \hlkwd{dsm}\hlstd{(count}\hlopt{~}\hlstd{chaparral} \hlopt{+}
                     \hlkwd{I}\hlstd{(chaparral}\hlopt{^}\hlnum{2}\hlstd{)} \hlopt{+}
                     \hlstd{elevation,}
               \hlkwc{observation.data}\hlstd{=obs_fall,} \hlkwc{segment.data}\hlstd{=segs,} \hlkwc{ddf.obj}\hlstd{=df_fall,}
               \hlkwc{transect}\hlstd{=}\hlstr{"point"}\hlstd{,} \hlkwc{family}\hlstd{=}\hlkwd{nb}\hlstd{())}

\hlcom{# check parameter estimates}
\hlkwd{summary}\hlstd{(ll_fall)}
\end{alltt}
\begin{verbatim}
## 
## Family: Negative Binomial(0.343) 
## Link function: log 
## 
## Formula:
## count ~ chaparral + I(chaparral^2) + elevation + offset(off.set)
## 
## Parametric coefficients:
##                Estimate Std. Error z value Pr(>|z|)    
## (Intercept)    -11.7170     0.1682 -69.656  < 2e-16 ***
## chaparral        1.4288     0.1906   7.496 6.57e-14 ***
## I(chaparral^2)  -0.3789     0.1255  -3.019  0.00254 ** 
## elevation       -0.2260     0.1478  -1.528  0.12639    
## ---
## Signif. codes:  0 '***' 0.001 '**' 0.01 '*' 0.05 '.' 0.1 ' ' 1
## 
## 
## R-sq.(adj) =  0.026   Deviance explained = 25.9%
## -REML = 270.86  Scale est. = 1         n = 307
\end{verbatim}
\end{kframe}
\end{knitrout}

We can then do the variance propagation via the \texttt{dsm\_varprop}
function and compare to the delta methd (implemented in the \texttt{dsm.var.gam}
function):
\begin{knitrout}
\definecolor{shadecolor}{rgb}{0.969, 0.969, 0.969}\color{fgcolor}\begin{kframe}
\begin{alltt}
\hlcom{# varprop}
\hlstd{pp_fall} \hlkwb{<-} \hlkwd{dsm_varprop}\hlstd{(ll_fall, pred)}
\hlstd{pp_fall}
\end{alltt}
\begin{verbatim}
## Summary of uncertainty in a density surface model calculated
##  by variance propagation.
## 
## Probability of detection in fitted model and variance model
##    chaparral Fitted.model Fitted.model.se Refitted.model
## 1 -0.8613374    0.3345638      0.04953141      0.3383015
## 2  0.1733685    0.2350608      0.02187221      0.2361407
## 3  2.0242947    0.1166973      0.01900894      0.1154638
## 
## Approximate asymptotic confidence interval:
##     2.5%     Mean    97.5% 
## 1625.431 2271.870 3175.400 
## (Using log-Normal approximation)
## 
## Detection function CV          : 0.0897 
## 
## Point estimate                 : 2271.87 
## Standard error                 : 390.9639 
## Coefficient of variation       : 0.1721
\end{verbatim}
\begin{alltt}
\hlcom{# delta method for comparison}
\hlstd{vg_fall} \hlkwb{<-} \hlkwd{dsm.var.gam}\hlstd{(ll_fall, pred,} \hlkwc{off.set}\hlstd{=pred}\hlopt{$}\hlstd{off.set)}
\hlstd{vg_fall}
\end{alltt}
\begin{verbatim}
## Summary of uncertainty in a density surface model calculated
##  analytically for GAM, with delta method
## 
## Approximate asymptotic confidence interval:
##     2.5%     Mean    97.5% 
## 1614.289 2270.436 3193.281 
## (Using log-Normal approximation)
## 
## Point estimate                 : 2270.436 
## CV of detection function       : 0.08974614 
## CV from GAM                    : 0.1506 
## Total standard error           : 398.1164 
## Total coefficient of variation : 0.1753
\end{verbatim}
\end{kframe}
\end{knitrout}

\subsection{Spring 2009}

Again fitting models:

\begin{knitrout}
\definecolor{shadecolor}{rgb}{0.969, 0.969, 0.969}\color{fgcolor}\begin{kframe}
\begin{alltt}
\hlcom{# fit detection function}
\hlstd{df_spring} \hlkwb{<-} \hlkwd{ds}\hlstd{(obs_spring,} \hlkwc{transect}\hlstd{=}\hlstr{"point"}\hlstd{,} \hlkwc{formula}\hlstd{=}\hlopt{~}\hlstd{forest)}
\end{alltt}

{\ttfamily\noindent\itshape\color{messagecolor}{\#\# Columns "{}distbegin"{} and "{}distend"{} in data: performing a binned analysis...}}

{\ttfamily\noindent\itshape\color{messagecolor}{\#\# Model contains covariate term(s): no adjustment terms will be included.}}

{\ttfamily\noindent\itshape\color{messagecolor}{\#\# Fitting half-normal key function}}

{\ttfamily\noindent\itshape\color{messagecolor}{\#\# AIC= 303.83}}

{\ttfamily\noindent\itshape\color{messagecolor}{\#\# No survey area information supplied, only estimating detection function.}}\begin{alltt}
\hlcom{# fit spatial model}
\hlstd{ll_spring} \hlkwb{<-} \hlkwd{dsm}\hlstd{(count}\hlopt{~}\hlstd{chaparral} \hlopt{+}
                       \hlkwd{I}\hlstd{(chaparral}\hlopt{^}\hlnum{2}\hlstd{)} \hlopt{+}
                       \hlstd{elevation} \hlopt{+}
                       \hlkwd{I}\hlstd{(elevation}\hlopt{^}\hlnum{2}\hlstd{),}
          \hlkwc{observation.data}\hlstd{=obs_spring,} \hlkwc{segment.data}\hlstd{=segs,} \hlkwc{ddf.obj}\hlstd{=df_spring,}
          \hlkwc{transect}\hlstd{=}\hlstr{"point"}\hlstd{,} \hlkwc{family}\hlstd{=}\hlkwd{nb}\hlstd{())}

\hlcom{# check parameter estimates}
\hlkwd{summary}\hlstd{(ll_spring)}
\end{alltt}
\begin{verbatim}
## 
## Family: Negative Binomial(0.672) 
## Link function: log 
## 
## Formula:
## count ~ chaparral + I(chaparral^2) + elevation + I(elevation^2) + 
##     offset(off.set)
## 
## Parametric coefficients:
##                 Estimate Std. Error z value Pr(>|z|)    
## (Intercept)    -11.43571    0.17886 -63.936  < 2e-16 ***
## chaparral        0.68155    0.15642   4.357 1.32e-05 ***
## I(chaparral^2)  -0.31515    0.10730  -2.937  0.00331 ** 
## elevation       -0.08065    0.15006  -0.537  0.59098    
## I(elevation^2)  -0.33865    0.15125  -2.239  0.02515 *  
## ---
## Signif. codes:  0 '***' 0.001 '**' 0.01 '*' 0.05 '.' 0.1 ' ' 1
## 
## 
## R-sq.(adj) =  0.0727   Deviance explained = 12.3%
## -REML = 276.52  Scale est. = 1         n = 307
\end{verbatim}
\end{kframe}
\end{knitrout}

Compare variance calculations:
\begin{knitrout}
\definecolor{shadecolor}{rgb}{0.969, 0.969, 0.969}\color{fgcolor}\begin{kframe}
\begin{alltt}
\hlcom{# varprop}
\hlstd{pp_spring} \hlkwb{<-} \hlkwd{dsm_varprop}\hlstd{(ll_spring, pred,} \hlkwc{trace}\hlstd{=}\hlnum{TRUE}\hlstd{)}
\hlstd{pp_spring}
\end{alltt}
\begin{verbatim}
## Summary of uncertainty in a density surface model calculated
##  by variance propagation.
## 
## Probability of detection in fitted model and variance model
##       forest Fitted.model Fitted.model.se Refitted.model
## 1 -0.4633652    0.2710221      0.02760400      0.2514265
## 2 -0.3424751    0.2604019      0.02555611      0.2468799
## 3  2.5274876    0.0933240      0.03014374      0.1569675
## 
## Approximate asymptotic confidence interval:
##     2.5%     Mean    97.5% 
## 1262.702 1684.184 2246.355 
## (Using log-Normal approximation)
## 
## Detection function CV          : 0.142 
## 
## Point estimate                 : 1684.184 
## Standard error                 : 248.8427 
## Coefficient of variation       : 0.1478
\end{verbatim}
\begin{alltt}
\hlcom{# delta method for comparison}
\hlstd{vg_spring} \hlkwb{<-} \hlkwd{dsm.var.gam}\hlstd{(ll_spring, pred,} \hlkwc{off.set}\hlstd{=pred}\hlopt{$}\hlstd{off.set)}
\hlstd{vg_spring}
\end{alltt}
\begin{verbatim}
## Summary of uncertainty in a density surface model calculated
##  analytically for GAM, with delta method
## 
## Approximate asymptotic confidence interval:
##     2.5%     Mean    97.5% 
## 1187.849 1697.548 2425.956 
## (Using log-Normal approximation)
## 
## Point estimate                 : 1697.548 
## CV of detection function       : 0.1420396 
## CV from GAM                    : 0.1165 
## Total standard error           : 311.8206 
## Total coefficient of variation : 0.1837
\end{verbatim}
\end{kframe}
\end{knitrout}

\section{Comparing to results in Sillet et al.}

Table \ref{tab:coef-table} compares the parameter estimates from
Sillet et al. (their Table 3) with the results from using the DSMs
described above.

To obtain uncertainty estimates of detection function parameters we
require the following function to recompute the Hessian for that model
component.

\begin{knitrout}
\definecolor{shadecolor}{rgb}{0.969, 0.969, 0.969}\color{fgcolor}\begin{kframe}
\begin{alltt}
\hlcom{# wee function to get a "corrected" detection function}
\hlcom{# note this only works for half-normal models}
\hlcom{# the function's argument object is the result from dsm_varprop}
\hlstd{fix_ddf_vp} \hlkwb{<-} \hlkwa{function}\hlstd{(}\hlkwc{object}\hlstd{)\{}
  \hlcom{# get the model from the varprop object}
  \hlstd{fix_ddf} \hlkwb{<-} \hlstd{object}\hlopt{$}\hlstd{old_model}\hlopt{$}\hlstd{ddf}
  \hlcom{# get parameters}
  \hlstd{which.names} \hlkwb{<-} \hlopt{!}\hlstd{(}\hlkwd{names}\hlstd{(}\hlkwd{coef}\hlstd{(object}\hlopt{$}\hlstd{refit))} \hlopt{%in%} \hlkwd{names}\hlstd{(}\hlkwd{coef}\hlstd{(object}\hlopt{$}\hlstd{old_model)))}
  \hlstd{newpar} \hlkwb{<-} \hlstd{fix_ddf}\hlopt{$}\hlstd{par} \hlopt{+} \hlkwd{coef}\hlstd{(object}\hlopt{$}\hlstd{refit)[which.names]}

  \hlcom{# setup a new fitting call}
  \hlstd{fix_call} \hlkwb{<-} \hlstd{fix_ddf}\hlopt{$}\hlstd{call}
  \hlstd{fix_ddf}\hlopt{$}\hlstd{control}\hlopt{$}\hlstd{initial} \hlkwb{<-} \hlkwd{list}\hlstd{()}
  \hlstd{fix_ddf}\hlopt{$}\hlstd{control}\hlopt{$}\hlstd{initial}\hlopt{$}\hlstd{scale} \hlkwb{<-} \hlstd{newpar}

  \hlcom{# ensure that the model is not fitted!}
  \hlcom{# Just want to calculate the Hessian!}
  \hlstd{fix_ddf}\hlopt{$}\hlstd{control}\hlopt{$}\hlstd{maxiter} \hlkwb{<-} \hlnum{0}
  \hlstd{fix_ddf}\hlopt{$}\hlstd{control}\hlopt{$}\hlstd{nrefits} \hlkwb{<-} \hlnum{0}
  \hlstd{fix_ddf}\hlopt{$}\hlstd{control}\hlopt{$}\hlstd{refit} \hlkwb{<-} \hlnum{FALSE}
  \hlstd{fix_ddf}\hlopt{$}\hlstd{control}\hlopt{$}\hlstd{nofit} \hlkwb{<-} \hlnum{TRUE}

  \hlcom{# run the model and return it}
  \hlstd{fix_ddf} \hlkwb{<-} \hlkwd{eval}\hlstd{(fix_call, fix_ddf)}
  \hlkwd{invisible}\hlstd{(fix_ddf)}
\hlstd{\}}


\hlcom{# corrected for fall}
\hlstd{vp_ddf_fall} \hlkwb{<-} \hlkwd{fix_ddf_vp}\hlstd{(pp_fall)}

\hlcom{# corrected for spring}
\hlstd{vp_ddf_spring} \hlkwb{<-} \hlkwd{fix_ddf_vp}\hlstd{(pp_spring)}
\end{alltt}
\end{kframe}
\end{knitrout}

\begin{table}
\begin{tabular}{lcccccc}
Submodel and &
\multicolumn{3}{c}{Fall 2008} &
\multicolumn{3}{c}{Spring 2009}\tabularnewline
~~coefficient &
Sillet et al. &
Pre-varprop &
Post-varprop &
Sillet et al. &
Pre-varprop &
Post-varprop\tabularnewline
\hline 
\hline 
Abundance &
 &
 &
 &
 &
 &
\tabularnewline
~Chaparral &
1.43 (0.229) &
1.43 (0.191) &
1.44 (0.209) &
0.67 (0.667) &
0.68 (0.156) &
0.67 (0.156)\tabularnewline
~Chaparral$^{2}$ &
-0.38 (0.115) &
-0.38 (0.126) &
-0.38 (0.126) &
-0.29 (-0.291) &
-0.32 (0.107) &
-0.29 (0.106)\tabularnewline
~Elevation &
-0.23 (0.146) &
-0.23 (0.148) &
-0.23 (0.148) &
-0.11 (-0.107) &
-0.08 (0.15) &
-0.11 (0.148)\tabularnewline
~Elevation$^{2}$ &
 &
 &
 &
-0.34 (0.148) &
-0.34 (0.151) &
-0.34 (0.15)\tabularnewline
Dispersion &
0.36 (0.0777) &
0.34  &
0.34  &
0.78 (0.239) &
0.71  &
0.71\tabularnewline
Detection &
 &
 &
 &
 &
 &
\tabularnewline
~Intercept &
4.68 (0.0658) &
4.67 (0.055) &
4.68 (0.055) &
4.63 (0.540) &
4.63 (0.051) &
4.64 (0.05)\tabularnewline
~Chaparral &
-0.20 (0.060) &
-0.19 (0.052) &
-0.2 (0.052) &
 &
 &
\tabularnewline
~Forest &
 &
 &
 &
-0.09 (0.043) &
-0.18 (0.06) &
-0.08 (0.062)\tabularnewline
\hline 
\end{tabular}

\caption{Comparison of parameter estimates from Sillet et al. to those using
DSM (before and after our variance propagation procedure). Note that
\texttt{mgcv} (and hence the \texttt{dsm} package) does not provide
uncertainty around the negative binomial parameter. Abundance intercept
parameters are not comparable and therefore omitted (due to differences
in how the offset is calculated in each model).\label{tab:coef-table}}
\end{table}

Table \ref{tab:abundance} compares the abundance estimates from from
Sillet et al. to the estimates here.

\begin{table}
\begin{tabular}{ccc}
 &
Fall 2008 &
Spring 2009\tabularnewline
\hline 
\hline 
Sillet et al. &
2267 (1613--3007) &
1705 (1212--2369)\tabularnewline
DSM &
2272 (1625-3175) &
1684 (1263-2246)\tabularnewline
\hline 
\end{tabular}

\caption{Comparison of abundance estimates between Sillet et al. and the above
analysis.\label{tab:abundance}}

\end{table}

\begin{figure}
\begin{knitrout}
\definecolor{shadecolor}{rgb}{0.969, 0.969, 0.969}\color{fgcolor}
\includegraphics[width=\maxwidth]{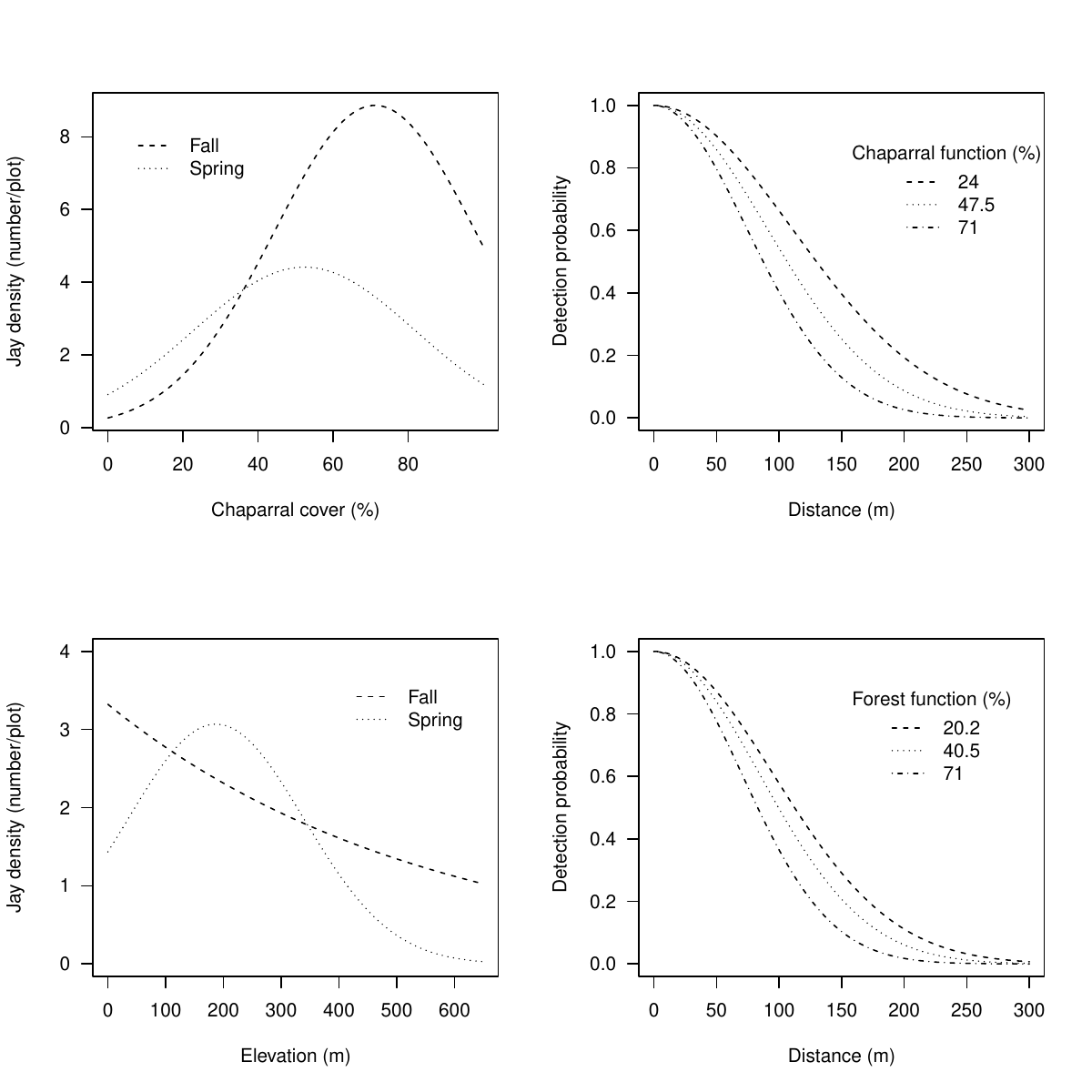} 

\end{knitrout}

\caption{Reproduction of Figure 4 from Sillet et al. using the models specified
above. Left side shows abundance per site as a function of the covariates
in each time period. Right side shows the detection function at selected
covariate levels for fall (top) and spring (bottom). These are indistinguishable
from those in the article.}

\end{figure}

\section{Unbalanced data}

Since the coverage of both covariate and geographical space is extremely
good, we decided to subsample the data to see what effect that had
on our models. We used the fall data only for this experiment and
randomly selected 100 points. We then removed points where where chaparral
cover was greater than the mean chaparral proportion (with mean calcualted
over all points).

\begin{knitrout}
\definecolor{shadecolor}{rgb}{0.969, 0.969, 0.969}\color{fgcolor}\begin{kframe}
\begin{alltt}
\hlcom{# reproducible subsample}
\hlkwd{set.seed}\hlstd{(}\hlnum{1889}\hlstd{)}

\hlcom{# take 100 sites at random}
\hlstd{sub_segs} \hlkwb{<-} \hlstd{segs[}\hlkwd{sample}\hlstd{(}\hlnum{1}\hlopt{:}\hlkwd{nrow}\hlstd{(segs),} \hlnum{100}\hlstd{), ]}
\hlcom{# only the observations at those sites are to be included}
\hlstd{sub_obs} \hlkwb{<-} \hlstd{obs_fall[obs_fall}\hlopt{$}\hlstd{Sample.Label} \hlopt{%in%} \hlstd{sub_segs}\hlopt{$}\hlstd{Sample.Label, ]}

\hlcom{# since the data were z-transformed, mean=0}
\hlcom{# so chaparral<=0 gives values at or lower than the mean}
\hlstd{sub_segs} \hlkwb{<-} \hlstd{sub_segs[sub_segs}\hlopt{$}\hlstd{chaparral}\hlopt{<=}\hlnum{0}\hlstd{, ]}
\hlstd{sub_obs} \hlkwb{<-} \hlstd{sub_obs[sub_obs}\hlopt{$}\hlstd{chaparral}\hlopt{<=}\hlnum{0}\hlstd{, ]}
\end{alltt}
\end{kframe}
\end{knitrout}
Having subsampled the data, we can then fit models:

\begin{knitrout}
\definecolor{shadecolor}{rgb}{0.969, 0.969, 0.969}\color{fgcolor}\begin{kframe}
\begin{alltt}
\hlcom{# fall detection function on subsampled data}
\hlstd{sub_df} \hlkwb{<-} \hlkwd{ds}\hlstd{(sub_obs,} \hlkwc{transect}\hlstd{=}\hlstr{"point"}\hlstd{,} \hlkwc{formula}\hlstd{=}\hlopt{~}\hlstd{chaparral)}
\end{alltt}

{\ttfamily\noindent\itshape\color{messagecolor}{\#\# Columns "{}distbegin"{} and "{}distend"{} in data: performing a binned analysis...}}

{\ttfamily\noindent\itshape\color{messagecolor}{\#\# Model contains covariate term(s): no adjustment terms will be included.}}

{\ttfamily\noindent\itshape\color{messagecolor}{\#\# Fitting half-normal key function}}

{\ttfamily\noindent\itshape\color{messagecolor}{\#\# AIC= 31.544}}

{\ttfamily\noindent\itshape\color{messagecolor}{\#\# No survey area information supplied, only estimating detection function.}}\begin{alltt}
\hlcom{# fall GAM}
\hlstd{sub_ll} \hlkwb{<-} \hlkwd{dsm}\hlstd{(count}\hlopt{~}\hlstd{chaparral} \hlopt{+}
              \hlkwd{I}\hlstd{(chaparral}\hlopt{^}\hlnum{2}\hlstd{)} \hlopt{+}
              \hlstd{elevation,}
              \hlkwc{observation.data}\hlstd{=sub_obs,} \hlkwc{segment.data}\hlstd{=sub_segs,} \hlkwc{ddf.obj}\hlstd{=sub_df,}
              \hlkwc{transect}\hlstd{=}\hlstr{"point"}\hlstd{,} \hlkwc{family}\hlstd{=}\hlkwd{nb}\hlstd{())}

\hlcom{# inspect coefficient estimates}
\hlkwd{summary}\hlstd{(sub_ll)}
\end{alltt}
\begin{verbatim}
## 
## Family: Negative Binomial(0.222) 
## Link function: log 
## 
## Formula:
## count ~ chaparral + I(chaparral^2) + elevation + offset(off.set)
## 
## Parametric coefficients:
##                Estimate Std. Error z value Pr(>|z|)    
## (Intercept)    -10.8052     1.6186  -6.676 2.46e-11 ***
## chaparral        3.7674     6.4030   0.588    0.556    
## I(chaparral^2)  -0.9836     5.1426  -0.191    0.848    
## elevation       -0.6331     0.4901  -1.292    0.196    
## ---
## Signif. codes:  0 '***' 0.001 '**' 0.01 '*' 0.05 '.' 0.1 ' ' 1
## 
## 
## R-sq.(adj) =  0.167   Deviance explained = 50.5%
## -REML = 26.603  Scale est. = 1         n = 65
\end{verbatim}
\end{kframe}
\end{knitrout}

Compare variance calculations:
\begin{knitrout}
\definecolor{shadecolor}{rgb}{0.969, 0.969, 0.969}\color{fgcolor}\begin{kframe}
\begin{alltt}
\hlcom{# variance propagation:}
\hlstd{sub_pp} \hlkwb{<-} \hlkwd{dsm_varprop}\hlstd{(sub_ll, pred)}
\hlstd{sub_pp}
\end{alltt}
\begin{verbatim}
## Summary of uncertainty in a density surface model calculated
##  by variance propagation.
## 
## Probability of detection in fitted model and variance model
##    chaparral Fitted.model Fitted.model.se Refitted.model
## 1 -0.9059225    0.7806223       0.5482305      0.6638134
## 2 -0.3043294    0.2713778       0.1688496      0.2792179
## 3 -0.1975517    0.1967457       0.1929185      0.2233795
## 
## Approximate asymptotic confidence interval:
##         2.5%         Mean        97.5% 
##     38.84062   2831.42634 206406.98577 
## (Using log-Normal approximation)
## 
## Detection function CV          : 0.7106 
## 
## Point estimate                 : 2831.426 
## Standard error                 : 30908.15 
## Coefficient of variation       : 10.9161
\end{verbatim}
\begin{alltt}
\hlcom{# delta method}
\hlstd{sub_vg} \hlkwb{<-} \hlkwd{dsm.var.gam}\hlstd{(sub_ll, pred,} \hlkwc{off.set}\hlstd{=pred}\hlopt{$}\hlstd{off.set)}
\hlstd{sub_vg}
\end{alltt}
\begin{verbatim}
## Summary of uncertainty in a density surface model calculated
##  analytically for GAM, with delta method
## 
## Approximate asymptotic confidence interval:
##        2.5%        Mean       97.5% 
##     208.649   26434.038 3348965.752 
## (Using log-Normal approximation)
## 
## Point estimate                 : 26434.04 
## CV of detection function       : 0.7106491 
## CV from GAM                    : 21.1064 
## Total standard error           : 558242.5 
## Total coefficient of variation : 21.1183
\end{verbatim}
\end{kframe}
\end{knitrout}

Get corrected detection function:
\begin{knitrout}
\definecolor{shadecolor}{rgb}{0.969, 0.969, 0.969}\color{fgcolor}\begin{kframe}
\begin{alltt}
\hlstd{sub_vp_ddf} \hlkwb{<-} \hlkwd{fix_ddf_vp}\hlstd{(sub_pp)}
\end{alltt}
\end{kframe}
\end{knitrout}

A summary of the abundance results before and after variance propagation
with those from the full data are given in Table \ref{tab:subsample-abund}.
Comparison of model coefficients is given in Table \ref{tab:subsample-par-table}.

\begin{table}
\begin{tabular}{ccc}
 &
Pre-varprop &
Post-varprop\tabularnewline
\hline 
\hline 
Full data &
2270 (1614-3193) &
2272 (1625-3175)\tabularnewline
Subsample &
26434 (209-3348966) &
2831 (39-206407)\tabularnewline
\hline 
\end{tabular}

\caption{Comparison of abundance estimates before and after subsampling.\label{tab:subsample-abund}}
\end{table}

\begin{table}
\begin{tabular}{lcccc}
Submodel and &
\multicolumn{2}{c}{Full data} &
\multicolumn{2}{c}{Subsample}\tabularnewline
~~coefficient &
Sillet et al &
DSM &
Pre-varprop &
Post-varprop\tabularnewline
\hline 
\hline 
Abundance &
 &
 &
 &
\tabularnewline
~Chaparral &
1.43 (0.229) &
1.44 (0.209) &
3.77 (6.403) &
2.56 (9.17)\tabularnewline
~Chaparral$^{2}$ &
-0.38 (0.115) &
-0.38 (0.126) &
-0.98 (5.143) &
-1.74 (6.541)\tabularnewline
~Elevation &
-0.23 (0.146) &
-0.23 (0.148) &
-0.63 (0.49) &
-0.63 (0.49)\tabularnewline
Dispersion &
0.36 (0.0777) &
0.34 &
0.22  &
0.22\tabularnewline
Detection &
 &
 &
 &
\tabularnewline
~Intercept &
4.68 (0.0658) &
4.68 (0.055) &
4.23 (0.955) &
4.39 (0.867)\tabularnewline
~Chaparral &
-0.20 (0.060) &
-0.2 (0.052) &
-1.61 (2.587) &
-1.14 (2.071)\tabularnewline
\hline 
\end{tabular}

\caption{Comparison of parameter estimates from the full fall data from Sillet
et al. and those using DSM, along with estimates for the subsampled
data before and after our variance propagation procedure. Note that
\texttt{mgcv} (and hence the \texttt{dsm} package) does not provide
uncertainty around the negative binomial parameter. Intercept parameters
are not comparable and therefore omitted (due to differences in how
the offset is calculated in each model). \label{tab:subsample-par-table}}
\end{table}


\title{Supplementary Material B: \\
Simulation study}
\author{Mark V. Bravington\\
David L. Miller\\
Sharon L. Hedley}
\maketitle

\section{Introduction}

To assess the properties of our proposed variance propagation method
we compared results to those from assuming independence (the ``delta
method'') and from a one-stage fully Bayesian approach using MCMC
to fit the model. We simulated data based on the island scrub jay
data used as an example in the paper. Our aim here is to investigate
the properties of our new method (versus the two other approaches)
in the situations where the number of detections decreases. We do
not concern ourselves with changing the sampling scheme or detection
function covariate for simplicity of presentation and analysis. We
also want to see what happens in the best possible situation, where
sampler coverage is good and the only ``problem'' is that detectability
decreases leading to a decrease in the number of animals observed.
As with any set of simulations, these should not be treated as a reliable
guide to performance in all conceivable situations.

\section{Simulation setup}

\begin{figure}
\begin{knitrout}
\definecolor{shadecolor}{rgb}{0.969, 0.969, 0.969}\color{fgcolor}
\includegraphics[width=\maxwidth]{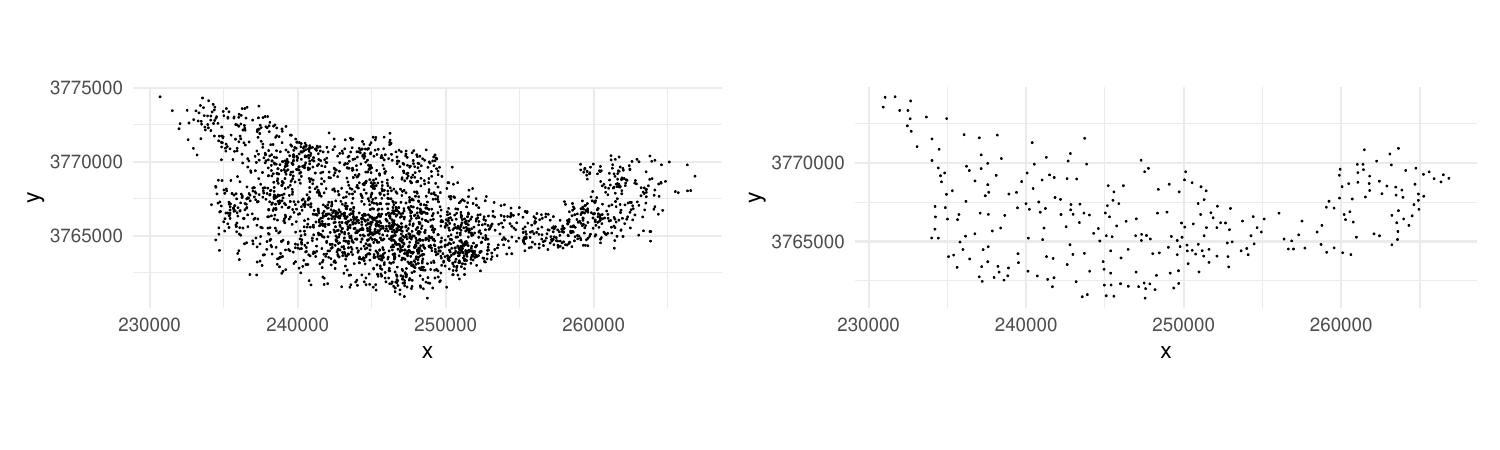} 

\end{knitrout}

\caption{Left: a simulated spatial distribution used in the simulation. Right:
sampler locations for the simulation study (as for the island scrub
jay example).\label{fig:samplers-distribution}}
\end{figure}

Our underlying setup was essentially the same our example for island
scrub jays for fall. The underlying density of animals was $\texttt{chaparrel}^{2}+\texttt{chaparrel}+\texttt{elevation}$
(normalized over the whole of Santa Cruz Island). We then sampled
each grid cell from the prediction grid using a Poisson distribution
with rate parameter a function of the underlying density, multiplied
by 2500 (our target total true abundance). This lead true animal locations
similar to those shown in the left panel of Figure \ref{fig:samplers-distribution}.
Point transect locations were left as in the example (Figure \ref{fig:samplers-distribution},
right panel).

Detection functions were varied but initially based on those resulting
from the fall model for island scrub jays: a half-normal detection
function with chaparral as a covariate with the parameter for chaparral
in the scale parameter fixed to be -0.2 throughout, the intercept
parameter for the scale was varied. To decide on these parameters
we effectively pursued a manual bisection on the intercept scale parameter:
first using parameter values from the example analysis as a guide
toe parameter values to use.

Since we only wished to find out about the general performance of
the procedures (rather than, say, calculate bias to a high precision),
we generated 200 realizations per simulation scenario.

\section{Model specification}

Three ``models'' were fitted to each data realization. In all cases,
first the largest 10\% of distances were discarded, this is not ideal
but prevented too many simulations failing due to models overfitting
to large distances (in practice one would review each model but this
is clearly not possible in a simulation setting). All models used
a bivariate thin plate regression spline with a basis size of 20 and
assumed a Poisson distribution for the response (for simplicity).

The first two models used the \texttt{dsm} package using REML estimation.
These models used the same fitted detection functions and spatial
models, however one model (``delta'') used analytic estimates assuming
independence between detection and spatial components whereas the
other (``varprop'') used our variance propagation procedure.

The third model is rather more involved: smoothers were setup using
the \texttt{mgcv::jagam} function and imported into a custom written
Nimble \parencite{devalpine2017programming,devalpine2020nimblemcmc}
code (essentially JAGS), which included additional code for simultaneous
fitting of the detection function. The spatial smoother was a thin
plate spline, split into two penalties by \texttt{mgcv::jagam} to
ensure priors were proper (for details see \cite{wood2016justanother}).

The full likelihood is the product of the detection function likelihood
($g(r_{i},\sigma_{i})/2\pi\int_{0}^{w}rg(r,\sigma_{i})\text{d}r$,
for detection $i$), the segment-level Poisson model and the priors
on the parameters to be estimated. Priors and intermediate calculations
were set as follows:
\begin{align*}
N_{j} & \sim\text{Poisson}(\mu_{j}),\\
\mu_{j} & =\nu_{j}\exp\left(\mathbf{X}\boldsymbol{\beta}\right),\\
\beta_{0} & \sim N(0,\tau_{\beta_{0}}),\\
\boldsymbol{\beta}_{1:19} & \sim\text{MVN}(\mathbf{0},\lambda_{1}\mathbf{S}_{1}+\lambda_{2}\mathbf{S}_{2}),\\
\lambda_{1},\lambda_{2} & =\Gamma(0.05,0.005),\\
\nu_{j} & =2\pi\int_{0}^{w}rg(r,\sigma_{j})\text{d}r,\\
g(r,\sigma) & =\exp\left(\frac{-r^{2}}{2\sigma^{2}}\right),\\
\sigma(\theta_{0},\theta_{\texttt{chap}},\texttt{chaparrel}) & =\exp\left(\theta_{0}+\theta_{\texttt{chap}}\texttt{chaparrel}\right),\\
\theta_{0} & \sim\text{N}(0,0.1)\\
\theta_{\texttt{chap}} & \sim\text{Unif}(-2,0).
\end{align*}
where $j$ indexes the segments and $i$ indexes the detections. $\sigma_{j}$
is shorthand for $\sigma(\theta_{0},\theta_{\texttt{chap}},\texttt{chaparrel}_{j})$,
the detection function scale parameter for segment $j$ and $\sigma_{i}$
is shorthand for $\sigma(\theta_{0},\theta_{\texttt{chap}},\texttt{chaparrel}_{i})$,
the scale parameter for detection $i$. MVN denotes a multivariate
normal distribution parameterized by its mean vector and precision
matrix and $\text{N}$ denotes a univariate normal distribution parameterized
by its mean and precision; $\Gamma$ denotes a gamma distribution
parameterized by its shape and rate (for consistency with Nimble code
in Supplementary Materials). $\tau_{\beta_{0}}$ is the precision
of the spatial model intercept (for details see \cite{wood2016justanother}).

We used a relatively infomative prior on $\theta_{\texttt{chap}}$
ensuring that this parameter was negative. This was reasonable as
we would expect a detectability covariate like foliage cover to be
negative \textit{a priori} (as it will decrease probability of detection).
Leaving priors too vague led to abundance estimates and associated
standard errors being orders of magnitude too large.

Developing the above required considerable trial-and-error and computational
resources. It is difficult to come up with priors that will work reliably
in a simulation setting while at the same time making the simulation
fair for all models. Our experience here was that the fully Bayesian
model is easy to outline but difficult to fully specify and fit (though
to some extent this issue is dataset dependent). Based on these test
runs and for managing computational load we selected a single chain
with 500,000 iterations, with 100,000 iterations of burn-in and thinning
at every 50th iteration.

All code is available as part of the Supplementary Materials.

\section{Model assessment\label{sec:Model-assessment}}

In order to assess how well our variance propagation method performs
under decreasing numbers of detections, we used a measure of ``model
self-confidence'' which takes into account both bias and variance
in the resulting estimates of total abundance. Assuming that the abundance
is log-normally distributed \parencite[e.g.,][Section 3.6]{buckland2001introduction},
we can use a model's estimated abundance and variance around that
estimate to evaluate the quantile of that distribution that the true
abundance lies in (i.e., calculate $\mathbb{P}(N_{\text{truth}}\leq\hat{N})$
for each model/simulation realization)\@. If our model performs perfectly
then we would expect that a resulting histogram of the quantiles would
be flat. Skew in the plot indicates over/under-estimation of abundance
(bias). A histogram with high values at either end but a ``gap''
in the middle indicates uncertainty is underestimated ($N_{\text{truth}}$
lies in the tails of the model distributions too often). A conservative
estimator will give a domed histogram around 0.5, indicating intervals
are too wide. The latter case seems desirable for conservation work,
but a flat histogram is also acceptable. A deficiency of these plots
is there is no indication of scale: one might have a very small variance
but the self-confidence plot may be slightly domed or U-shaped, for
this reason we also calculated the average CV for the given parameter/model
combination.

For our variance propagation method and the fully Bayesian approach,
we could have used posterior draws to find the quantile of $N_{\text{truth}}$
in the posterior distribution, though this would not be possible for
the delta method approach. We did test this for a subset of the simulations
and found results were similar.

\section{Results}

\begin{figure}
\begin{knitrout}
\definecolor{shadecolor}{rgb}{0.969, 0.969, 0.969}\color{fgcolor}
\includegraphics[width=\maxwidth]{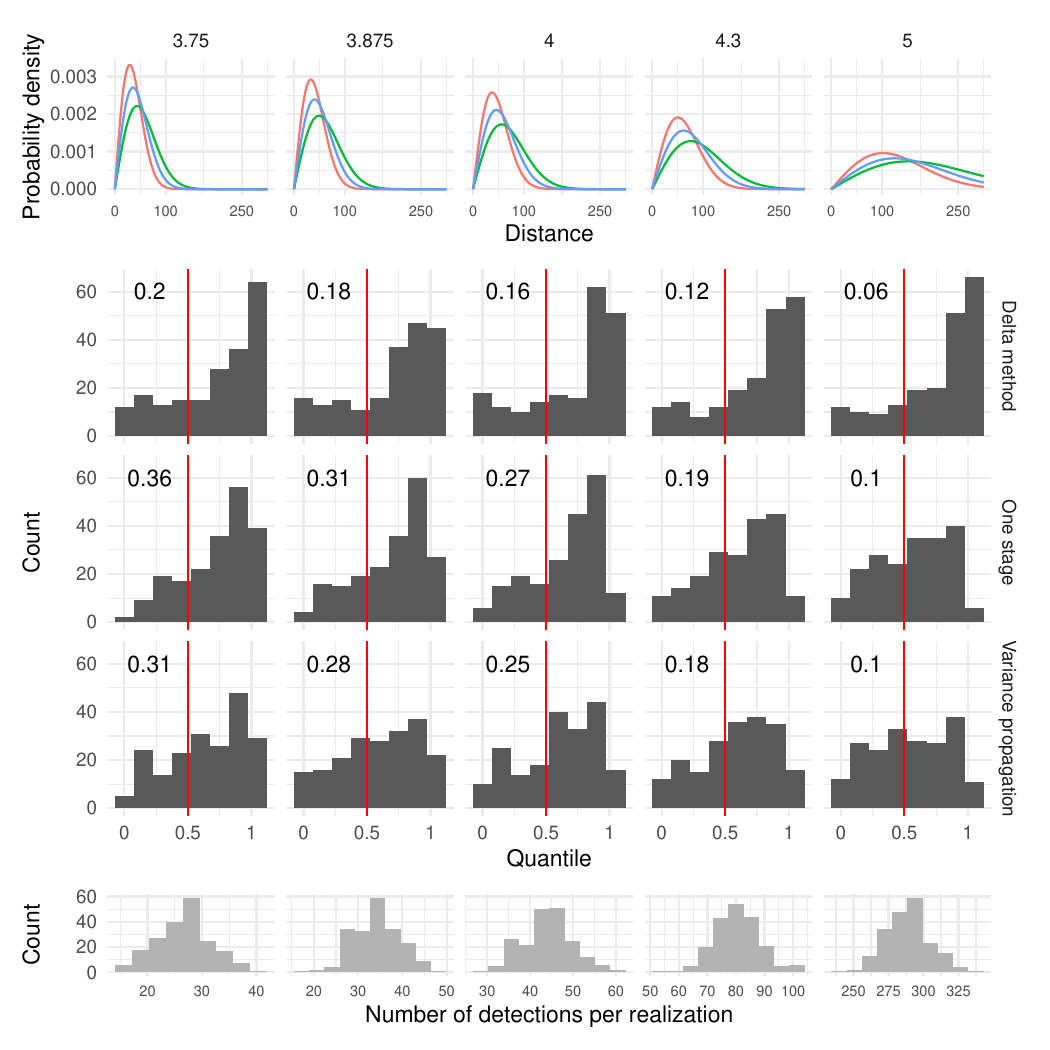} 

\end{knitrout}

\caption{Top row: detection functions evaluates at different levels of chaparral
(green=24, blue=47.5, red=71), top panel text gives the value of $\theta_{0}$
used in the simulation, $\theta_{\texttt{chap}}$ was fixed to -0.2
for all simulations. Middle 3 rows: ``Self-confidence'' histograms
showing the quantile that the true abundance lies in for each model/simulation
combination. CVs (averaged over the simulations) are printed in the
top left corner of each facet. Bottom row: distribution of number
of detections.\label{fig:Self-confidence-histograms}}
\end{figure}

Figure \ref{fig:Self-confidence-histograms} and Table \ref{tab:simres}
summarize the results from our simulations. As we might expect, we
see that performance of all methods degrades with decreasing sample
size of number of detections (as mediated by the change in the detection
function). The delta method is most biased in all scenarios as the
model is not able to correct point estimates of detection function
parameters nor estimate the covariance between the spatial model and
detection function. Both our variance propagation and the one stage
method perform well in the easier situations where sample size is
large, with performance degrading towards bias as sample size decreases
(working down Table \ref{tab:simres} or left to right in Figure \ref{fig:Self-confidence-histograms}).
The delta method gives a much smaller CV on average than either other
method, with the our variance propagation method and the one stage
approach giving similar results (top left corner of facets in Figure
\ref{fig:Self-confidence-histograms}). Coverage of the one stage
and variance propagation methods appears comparable throughout and
close to nomial levels at the three higher sample sizes (Table \ref{tab:simres},
last column). Surprisingly, our variance propagation method actually
performs better in terms of bias than the one-stage method in these
simulations, despite the approximation involved in ours (Table \ref{tab:simres},
third column). Presumably, that somehow reflects the choices we made
for priors in the one-stage method. However, as noted earlier, prior
choice can be a genuine conundrum in one-stage DSMs not just for simulations
but also for real-data applications, and one which our method circumvents.

\begin{table}

\begin{tabular}{lrrrrrrr}
\toprule
Model & $\theta_0$ & Bias($\hat{N}$) & se(Bias($\hat{N}$)) & $\text{Var}(\hat{N})$ & $\text{Var}(N)$ & MSE & Coverage\\
\midrule
Delta method & 5.000 & 176.103 & 233.417 & 54776.25 & 2306.094 & 85259.71 & 0.635\\
One stage & 5.000 & 44.657 & 237.591 & 56204.32 & 2306.094 & 57476.17 & 0.965\\
Variance propagation & 5.000 & 22.889 & 261.906 & 68181.10 & 2306.094 & 68845.74 & 0.930\\
\addlinespace
Delta method & 4.300 & 305.214 & 419.813 & 176146.26 & 2090.242 & 237692.74 & 0.680\\
One stage & 4.300 & 155.713 & 457.663 & 210850.35 & 2090.242 & 220701.52 & 0.940\\
Variance propagation & 4.300 & 116.430 & 469.337 & 221001.83 & 2090.242 & 225429.51 & 0.920\\
\addlinespace
Delta method & 4.000 & 350.359 & 565.371 & 315493.80 & 2616.400 & 391315.46 & 0.690\\
One stage & 4.000 & 277.038 & 601.148 & 356283.45 & 2616.400 & 424801.21 & 0.945\\
Variance propagation & 4.000 & 144.868 & 629.705 & 394071.33 & 2616.400 & 404599.59 & 0.925\\
\addlinespace
Delta method & 3.875 & 318.274 & 627.819 & 395875.89 & 2482.191 & 462374.56 & 0.760\\
One stage & 3.875 & 433.910 & 701.953 & 496647.78 & 2482.191 & 599028.97 & 0.880\\
Variance propagation & 3.875 & 107.519 & 777.841 & 605445.53 & 2482.191 & 606090.59 & 0.870\\
\addlinespace
Delta method & 3.750 & 384.447 & 726.423 & 533489.83 & 2627.598 & 658710.47 & 0.690\\
One stage & 3.750 & 531.775 & 771.133 & 602938.84 & 2627.598 & 828406.24 & 0.840\\
Variance propagation & 3.750 & 247.203 & 846.123 & 721152.69 & 2627.598 & 761978.21 & 0.870\\
\bottomrule
\end{tabular}

\caption{Simulation results showing (per model-parameter set combination):
median bias and its standard error, variance of the abundance estimate
(which includes variance in the true $N$, which changed per simulated
dataset), variance of true abundance (to give comparison to the previous
column), mean squared error (MSE), and coverage (proportion of simulations
where the true abundance was within the 95\% interval, constructed
via a $\log$-normal approximation).\label{tab:simres}}
\end{table}
\printbibliography


\title{Supplementary Material C: \\
Asymptotic properties of the approximation used for variance propagation}
\author{Mark V. Bravington\\
David L. Miller\\
Sharon L. Hedley}

\maketitle
The purpose of this Supplementary Material is to show that our approximation
does not worsen the asymptotic performance when estimating smoother
coefficients $\beta$ or the variance\slash smoothing-related hyperparameters
$\omega$, compared to an \textquotedbl idealized REML\textquotedbl{}
alternative that would (somehow) still use Laplace Approximation to
estimate $\omega$ but without the approximations related to the detection
probabilities. We assume that the problem is sufficiently regular
for standard asymptotics to apply (for example: parameter estimates
not on the boundary; priors that support the truth, etc; see e.g.
\cite{Pfanzagl1973asympedgex} and \cite{Weng2010bayesedgex}), and
in particular so that the use of a smoother can be justified in the
first place (\cite{wood2016smooths-mgcv}).

A further complication arises in the detection function\slash density
surface setting. For the detection function parameters $\theta$,
the number of sightings $n=\sum y_{i}$ constitutes the sample size,
and is an ancillary statistic that in itself carries no information
about $\theta$ (though it does affect the precision). However, $n$
itself is a random variable that is directly affected by $\beta$
in the density surface (in particular, by the intercept $\beta_{0}$).
A more relevant asymptotic quantity for the density surface would
be total survey effort $A$, corresponding to overall area searched.
Here we assume that $A$ would be increased by replicating the segments.
In practice, of course, available survey effort tends to be used not
to repeat segments but rather to infill between existing ones, so
that the design points $X_{i}$ become increasingly dense but are
not replicated; however, asymptotic arguments are much simpler in
the repeated case.

Our goal here is not statistical rigour per se, but rather to highlight
the comparative importance of terms included and excluded in our approximation.
Therefore, we make a number of simplifications at the loss of generality,
in particular by treating the detection function parameter as if it
were one-dimensional, and assuming that the count data are Poisson;
those assumptions do not affect the flow of the logic, but do considerably
simplify the algebra. The additivity property of Poisson distributions
means that increasing the \textquotedbl sample size\textquotedbl{}
(by repeating segments) has the same statistical effect as simply
increasing the search effort within each segment. Therefore we can
keep the number of segments fixed in the algebra, and write instead

\begin{gather*}
\mu_{i}\triangleq\mathbb{E}\left[Y_{i}\right]\\
\log\mathbb{E}\left[Y_{i}\right]=\log A+q_{i}\left(\theta\right)+X_{i}^{\top}\beta
\end{gather*}
and deal with asymptotics as $A\rightarrow\infty$. The number of
sightings $n$, on which inferences about detection-function parameters
$\theta$ are based, is $O_{p}\left(A\right)$, and for any $i$ we
have $\mathbb{E}\left[Y_{i}\right]=O_{p}\left(A\right)$ and $\mathbb{V}\left[Y_{i}\right]=O_{p}\left(A\right)$.

Below we use standard formulae for Poisson distributions, in this
form
\begin{gather*}
\log f\left(y|\mu_{i}\right)=C+y_{i}\log\mu_{i}-\mu_{i}\\
\frac{d\mu_{i}}{d\beta}=X_{i}\mu_{i}\\
\frac{d\mu_{i}}{d\delta}=\frac{d\mu_{i}}{d\theta}=q_{i}'\mu_{i}
\end{gather*}

where the \textquotedbl constant\textquotedbl{} $C$ varies from
line to line.

We further assume that detection function estimation takes place in
a Bayesian context, leading to a detection posterior $f_{n}\left(\theta\right)$.
(In practice, MLE will often be used, implicitly corresponding to
a weak prior.) We assume an Edgeworth-like expansion for $f_{n}$,
centred on the detection posterior mode $\hat{\theta}$. While such
expansions are routine in MLE settings, the Bayesian setting entails
more algebra and some extra regularity conditions, basically to ensure
that the influence of the Bayesian prior is asymptotically negligible;
see eg \textcite{Weng2010bayesedgex}. The expansion nevertheless
has the familiar form, given in eqn~(52) of that paper. In the one-dimensional
case used here for simplicity, it can be written as 
\begin{gather*}
f_{n}\left(\theta\right)=\sqrt{v}\times\varphi\left(\frac{\delta}{\sqrt{v}}\right)\times\left(1+O\left(n^{-1/2}\right)\text{poly}_{2}\left(\frac{\delta}{\sqrt{v}}\right)\right)\\
\implies\log f_{n}\left(\theta\right)=\tfrac{1}{2}\log v-\frac{\delta^{2}}{2v}+O\left(n^{-1/2}\right)\text{poly}_{2}\left(\frac{\delta}{\sqrt{v}}\right)
\end{gather*}

where $\delta=\theta-\hat{\theta}$, $v=\mathbb{V}\left[\theta|s\right]=O\left(n^{-1}\right)$,
$\varphi\left(\cdot\right)$ is a standardized Gaussian PDF, and $\text{poly}_{2}\left(x\right)$
is a polynomial with leading term in $x^{2}$ formed from Hermite
polynomials; the term in $x^{1}$ must vanish, because the expansion
is centred on the mode $\delta=0$ where $d\log f_{n}\left(\theta\right)/d\theta=0$.
Replacing $v$ by an estimate $\hat{v}$ (the detection posterior
variance) does not affect the order of asymptotic expansion that we
will require. Standard asymptotic results are that $\delta=O_{p}\left(n^{-1/2}\right)$,
$v=O_{p}\left(n^{-1}\right)$, $\delta/\sqrt{v}=O_{p}\left(1\right)$.
Because $n=O_{p}\left(A\right)$, we also have e.g. $\delta^{2}=O_{p}\left(A^{-1}\right)$,
etc.

Conditional on smoothing parameters $\omega$ and detection data $s$,
the joint log-density of the counts $y$, the smoother coefficients
$\beta$, and the detection parameters $\theta$ is

\begin{gather}
\Lambda\triangleq\log f\left(y,\beta,\theta|s\right)\nonumber \\
=C+\sum\left(y_{i}\log\mu_{i}-\mu_{i}\right)-\frac{1}{2}\beta^{\top}W\beta+\log f_{n}\left(\theta\right)\nonumber \\
=\sum y_{i}\log A+\sum y_{i}\left(q_{i}\left(\theta\right)+\beta^{\top}X_{i}\right)-A\sum\exp\left(q_{i}\left(\theta\right)+\beta^{\top}X_{i}\right)-\frac{1}{2}\beta^{\top}W\beta+\log f_{n}\left(\theta\right)\nonumber \\
=C+\sum y_{i}\left(q_{i}+\beta^{\top}X_{i}\right)-A\sum\exp\left(q_{i}+\beta^{\top}X_{i}\right)-\frac{1}{2}\beta^{\top}W\beta-\frac{\delta^{2}}{2v}+O\left(n^{-1/2}\right)\text{poly}_{2}\left(\frac{\delta}{\sqrt{v}}\right)\label{eq:Lambda}
\end{gather}

where $W$ is the prior precision matrix for $\beta$. Our approximation
$\tilde{\Lambda}$ omits the higher-order terms in $f_{n}$ and uses
a linear approximation to $q$, to give

\begin{gather}
\tilde{\mu}_{i}=A\exp\left(\hat{q}_{i}+\hat{q}'_{i}\delta+\beta^{\top}X_{i}\right)\nonumber \\
\tilde{\Lambda}=C+\sum\left(y_{i}\log\tilde{\mu}_{i}-\tilde{\mu}_{i}\right)-A\sum\exp\left(\hat{q}_{i}+\hat{q}'_{i}\delta+\beta^{\top}X_{i}\right)-\frac{1}{2}\beta^{\top}W\beta-\frac{\delta^{2}}{2v}\label{eq:Lambda-tilde}
\end{gather}

Since $q_{i}=\hat{q}_{i}+\hat{q}'_{i}\delta+\tfrac{1}{2}\hat{q}''_{i}\delta^{2}+O\left(\delta^{3}\right)$,
we have
\begin{gather*}
\exp\left(\hat{q}_{i}+\hat{q}'_{i}\delta+\beta^{\top}X_{i}\right)=\exp\left(q_{i}+\beta^{\top}X_{i}-\tfrac{1}{2}\hat{q}''_{i}\delta^{2}+O\left(\delta^{3}\right)\right)\\
=\exp\left(q_{i}+\beta^{\top}X_{i}\right)\exp\left(-\tfrac{1}{2}\hat{q}''_{i}\delta^{2}+O\left(\delta^{3}\right)\right)\\
=\exp\left(q_{i}+\beta^{\top}X_{i}\right)\left(1-\tfrac{1}{2}\hat{q}''_{i}\delta^{2}+O\left(\delta^{3}\right)+O\left(\delta^{4}\right)\right)\\
=\exp\left(q_{i}+\beta^{\top}X_{i}\right)+\left(-\tfrac{1}{2}\hat{q}''_{i}\delta^{2}+O\left(\delta^{3}\right)\right)\exp\left(q_{i}+\beta^{\top}X_{i}\right)\\
\implies\tilde{\mu}_{i}=\mu_{i}\left(1-\tfrac{1}{2}\hat{q}''_{i}\delta^{2}+O\left(\delta^{3}\right)\right)
\end{gather*}

and thus we can write eqn~(\ref{eq:Lambda-tilde}) as

\begin{gather*}
\tilde{\Lambda}=C+\sum y_{i}\left(q_{i}-\tfrac{1}{2}\hat{q}''_{i}\delta^{2}+O\left(\delta^{3}\right)+\beta^{\top}X_{i}\right)-A\sum\exp\left(q_{i}-\tfrac{1}{2}\hat{q}''_{i}\delta^{2}+O\left(\delta^{3}\right)+\beta^{\top}X_{i}\right)\\
-\frac{1}{2}\beta^{\top}W\beta-\frac{\delta^{2}}{2v}-O\left(n^{-1/2}\right)\text{poly}_{2}\left(\frac{\delta}{\sqrt{v}}\right)\\
=C+\sum\left(y_{i}\log\mu_{i}-\mu_{i}\right)+\sum y_{i}\left(-\tfrac{1}{2}\hat{q}''_{i}\delta^{2}+O\left(\delta^{3}\right)\right)-A\sum\left(-\tfrac{1}{2}\hat{q}''_{i}\delta^{2}+O\left(\delta^{3}\right)\right)\exp\left(q_{i}+\beta^{\top}X_{i}\right)-O\left(n^{-1/2}\right)\text{poly}_{2}\left(\frac{\delta}{\sqrt{v}}\right)\\
=\Lambda+\sum\left(y_{i}-\mu_{i}\right)\left(-\tfrac{1}{2}\hat{q}''_{i}\delta^{2}+O\left(\delta^{3}\right)\right)-O\left(n^{-1/2}\right)\text{poly}_{2}\left(\frac{\delta}{\sqrt{v}}\right)
\end{gather*}

Inference is determined not by $\Lambda$ and $\Lambda'$ directly,
but rather for their derivatives WRTO $\beta$ and $\delta$, for
which we have:
\begin{gather}
\frac{d\tilde{\Lambda}}{d\beta}=\frac{d\Lambda}{d\beta}-\sum X_{i}\mu_{i}\left(-\tfrac{1}{2}\hat{q}''_{i}\delta^{2}+O\left(\delta^{3}\right)\right)\label{eq:dlt-db}\\
\frac{d\tilde{\Lambda}}{d\delta}=\frac{d\Lambda}{d\delta}-\sum q_{i}'\mu_{i}\left(-\tfrac{1}{2}\hat{q}''_{i}\delta^{2}+O\left(\delta^{3}\right)\right)+\sum\left(y_{i}-\mu_{i}\right)\left(-\hat{q}''_{i}\delta+O\left(\delta^{2}\right)\right)-\frac{1}{\sqrt{v}}O\left(n^{-1/2}\right)\text{poly}_{1}\left(\frac{\delta}{\sqrt{v}}\right)\label{eq:dlt-dd}
\end{gather}

The leading term in

\begin{gather*}
\frac{d\Lambda}{d\beta}=\sum X_{i}\left(y_{i}-\mu_{i}\right)-\beta^{\top}W
\end{gather*}
is a random variable with mean 0 and variance $O\left(A\right)$,
and thus has typical magnitude $O\left(A^{1/2}\right)$. This dominates
the term $\beta^{\top}W$ which is $O\left(1\right)$, so we can write
\begin{gather*}
\frac{d\Lambda}{d\beta}=O_{p}\left(A^{1/2}\right)
\end{gather*}
and, by comparison, the second term in eqn~(\ref{eq:dlt-db}) is
$O_{p}\left(A\right)\times O_{p}\left(A^{-1}\right)=O_{p}\left(1\right)$
where the factor $O_{p}\left(A\right)$ comes from $\mu_{i}$, and
the factor $O_{p}\left(A^{-1}\right)$ is the asympotic magnitude
of $\delta^{2}$. Consequently, $d\Lambda/d\beta=O_{p}\left(A^{1/2}\right)$
and our approximation $d\tilde{\Lambda}/d\beta$ differs from it by
a term only of $O_{p}\left(1\right)$. \\
\begin{gather}
\frac{d\Lambda}{d\beta}=O_{p}\left(A^{1/2}\right)\nonumber \\
\frac{d\tilde{\Lambda}}{d\beta}=\frac{d\Lambda}{d\beta}+O_{p}\left(1\right)\nonumber \\
=\frac{d\Lambda}{d\beta}\left(1+O_{p}\left(A^{-1/2}\right)\right)\label{eq:dlt-db-1}
\end{gather}

For $\delta$, we have

\begin{gather*}
\frac{d\Lambda}{d\delta}=O_{p}\left(A^{1/2}\right)\\
\sum q_{i}'\mu_{i}\left(-\tfrac{1}{2}\hat{q}''_{i}\delta^{2}+O\left(\delta^{3}\right)\right)=O\left(A\right)\times O\left(A^{-1}\right)=O_{p}\left(1\right)\\
\sum\left(y_{i}-\mu_{i}\right)\left(-\hat{q}''_{i}\delta+O\left(\delta^{2}\right)\right)=O_{p}\left(A^{1/2}\right)\times O_{p}\left(A^{-1/2}\right)=O_{p}\left(1\right)\\
\frac{1}{\sqrt{v}}O\left(n^{-1/2}\right)\text{poly}_{1}\left(\frac{\delta}{\sqrt{v}}\right)=O_{p}\left(A^{-1}\right)^{-1/2}\times O_{p}\left(A^{-1/2}\right)\times O_{p}\left(1\right)=O_{p}\left(1\right)
\end{gather*}

so that again
\begin{gather}
\frac{d\Lambda}{d\delta}=O_{p}\left(A^{1/2}\right)\nonumber \\
\frac{d\tilde{\Lambda}}{d\delta}=\frac{d\Lambda}{d\delta}+O_{p}\left(1\right)\nonumber \\
=\frac{d\Lambda}{d\delta}\left(1+O_{p}\left(A^{-1/2}\right)\right)\label{eq:dlt-dd-1}
\end{gather}

Writing $\psi\triangleq\left(\beta,\delta\right)$ and extending the
computations to second order shows that

\begin{gather}
\frac{d^{2}\Lambda}{d\psi^{2}}=O_{p}\left(A\right)\nonumber \\
\frac{d^{2}\tilde{\Lambda}}{d\psi^{2}}=\frac{d^{2}\Lambda}{d\psi^{2}}+O_{p}\left(A^{1/2}\right)\label{eq:approx-derivs-oa}
\end{gather}

Now, writing $\Lambda\left(\psi;y\right)$ to emphasize the dependence
on count data, the ideal posterior mode $\hat{\psi}$ satisfies 
\begin{gather*}
\Lambda'\left(\hat{\psi};y\right)\triangleq\left.\frac{d\Lambda}{d\psi}\right\vert _{\hat{\psi}}=0
\end{gather*}
whereas our approximation finds $\tilde{\psi}$ such that
\begin{gather*}
\tilde{\Lambda}'\left(\tilde{\psi};y\right)\triangleq\left.\frac{d\tilde{\Lambda}}{d\psi}\right\vert _{\tilde{\psi}}=0
\end{gather*}

The equations leading up to (\ref{eq:approx-derivs-oa}) show that
$\tilde{\Lambda}$ is an asymptotically small perturbation of $\Lambda$
of order $O\left(A^{-1/2}\right)$ compared to $\Lambda$ itself,
and that the same holds for the derivatives. Hence it is natural to
expect that the difference in the roots $\tilde{\psi}-\hat{\psi}$
should also be asymptotically small compared to $\hat{\psi}$. The
fact that $\tilde{\psi}$ and $\hat{\psi}$ are defined implicitly
rather than explicitly makes a formal proof cumbersome, but here we
sketch a approach based on Taylor expansions around the true value
$\psi_{0}$. We take $\psi_{0}=0$ WLOG, and for convenience write
$\Lambda_{0}\triangleq\Lambda\left(0;y\right)$ and replace tensor
expressions such as $\psi^{\top}\Lambda''_{0}$ by their scalar equivalents
such as $\psi\Lambda''_{0}$, etc, so that
\[
\Lambda'\left(\psi\right)=\Lambda'_{0}+\psi\Lambda''_{0}+O\left(\psi^{2}\right)
\]

Since $\hat{\psi}=O_{p}\left(A^{-1/2}\right)$ by standard arguments,
the terms in $\psi^{2}$ and higher become asymptotically negligible
near the root $\hat{\psi}$ compared to terms in $\psi^{0}$ and $\psi^{1}$
, so that
\begin{gather*}
\Lambda'\left(\hat{\psi}\right)=0\implies\hat{\psi}=-\left(\frac{\Lambda'_{0}}{\Lambda''_{0}}\right)+O_{p}\left(A^{-1}\right)
\end{gather*}

A formal development would entail multivariate series inversion; see
e.g. \cite{BNC1989asymp}, section 6.10.

For the approximate solution $\tilde{\psi}$, the analogous expression
is

\begin{gather*}
\tilde{\psi}=-\left(\frac{\tilde{\Lambda}'_{0}}{\tilde{\Lambda}''_{0}}\right)+O_{p}\left(A^{-1}\right)
\end{gather*}

Finally, we can substitute from eqns~(\ref{eq:dlt-db-1}) and (\ref{eq:dlt-dd-1})
so that 
\begin{gather*}
\tilde{\psi}=-\left(\frac{\Lambda'_{0}\left(1+O_{p}\left(A^{-1/2}\right)\right)}{\Lambda''_{0}\left(1+O_{p}\left(A^{-1/2}\right)\right)}\right)+O_{p}\left(A^{-1}\right)\\
=-\left(\frac{\Lambda'_{0}}{\Lambda''_{0}}\right)\left(1+O_{p}\left(A^{-1/2}\right)\right)+O_{p}\left(A^{-1}\right)\\
=\hat{\psi}+O_{p}\left(A^{-1}\right)
\end{gather*}

The point here is that the posterior mode of $\tilde{\Lambda}$ is
close to the posterior mode of the ideal $\Lambda$, i.e. to order
$O\left(A^{-1}\right)$, and a similar argument applies to the Hessian.
Since the Laplace Approximation on which REML estimates of hyperparameters
involves only the mode and the Hessian, the same accuracy should apply
to hyperparameter estimates, and then to inference about $\beta$
conditional on hyperparameters. And since Laplace Approximations themselves
are generally valid to $O\left(n^{-1}\right)$ (\cite{tierney1986postexplap}),
there is no loss of asymptotic order from the approximations we propose.
In contrast, though, the derivations above show that naive estimates
of $\beta$ and $\omega$ based on treating the detection posterior
mode $\hat{\theta}$ as exact, i.e. fixing $\delta=0$ and neglecting
even the first-order terms in $\delta$ which our approximation does
include, would be of lower asymptotic accuracy.

\printbibliography